\documentclass[a4paper,12pt]{article}
\usepackage{graphicx}

\input{tcilatex}

\begin{document}

\setcounter{page}{0} \topmargin 0pt \oddsidemargin 5mm \renewcommand{%
\thefootnote}{\fnsymbol{footnote}} \newpage \setcounter{page}{0} 
\begin{titlepage}
\begin{flushright}
Berlin Sfb288 Preprint 476 \\
US-FT/13-00\\
hep-th/0008044 \\
revised version
\end{flushright}
\vspace{0.2cm}
\begin{center}
{\Large {\bf Identifying  the Operator Content,} }

{\Large {\bf the  Homogeneous
Sine-Gordon models} }

\vspace{0.8cm}
{\large  O.A.~Castro-Alvaredo$^\sharp$ and  A.~Fring$\,^\star$ }

\vspace{0.2cm}
{\em $^\sharp$Departamento de F\'\i sica de Part\'\i culas, Facultad de F\'\i sica \\
Universidad de Santiago de Compostela\\
E-15706 Santiago de Compostela, Spain\\
\smallskip
$^\star$Institut f\"ur Theoretische Physik, 
Freie Universit\"at Berlin\\ 
Arnimallee 14, D-14195 Berlin, Germany }
\end{center}
\vspace{0.5cm}
 
\renewcommand{\thefootnote}{\arabic{footnote}}
\setcounter{footnote}{0}

\begin{abstract}
We address the general question of how to reconstruct the field content of a quantum
field theory from a given scattering theory in the context of the form factor program. 
For the $SU(3)_2$-homogeneous Sine-Gordon model we construct systematically 
all $n$-particle form factors for a huge class of operators
 in terms of general determinant formulae. We investigate how 
different  operators are interrelated by the momentum space cluster property.
Finally we compute several
two-point correlation functions and carry out the ultraviolet limit in order to
identify each operator  with its corresponding partner in the underlying conformal field
theory.
\par\noindent
PACS numbers: 11.10Kk, 11.55.Ds, 05.70.Jk, 05.30.-d, 64.60.Fr, 11.30.Er
\end{abstract}
\vfill{ \hspace*{-9mm}
\begin{tabular}{l}
\rule{6 cm}{0.05 mm}\\
Castro@fpaxp1.usc.es\\
Fring@physik.fu-berlin.de
\end{tabular}}
\end{titlepage}
\newpage

\section{Introduction}

The central concepts of relativistic quantum field theory, like Einstein
causality and Poincar\'{e} covariance, are captured in local field equations
and commutation relations. As a matter of fact local quantum physics
(algebraic quantum field theory) \cite{Haag} takes the collection of all
operators localized in a particular region, which generate a von Neumann
algebra, as its very starting point (for recent reviews see e.g. \cite{Buch}%
).

On the other hand, in the formulation of a quantum field theory one may
alternatively start from a particle picture and investigate the
corresponding scattering theories. In particular for 1+1 dimensional
integrable quantum field theories this latter approach has proved to be
impressively successful. As its most powerful tool one exploits here the
bootstrap principle \cite{boot}, which allows to write down exact scattering
matrices. Ignoring subtleties of non-asymptotic states, it is essentially
possible to obtain the latter picture from the former by means of the
LSZ-reduction formalism \cite{LSZ}. However, the question of how to
reconstruct at least part of the field content from the scattering theory is
in general still an outstanding issue. Recently a link between scattering
theory and local interacting fields in terms of polarization-free generators
has been developed \cite{Bert}. Unfortunately, they involve subtle domain
properties and are therefore objects which concretely can only be handled
with great difficulties.

In the context of 1+1 dimensional integrable quantum field theories the
identification of the operators is based on the assumption, dating back to
the initial papers \cite{Kar}, that each solution to the form factor
consistency equations \cite{Kar,Smir,Zamocorr,BFKZ} corresponds to a
particular local operator. Based on this numerous authors \cite
{Kar,Smir,Zamocorr,BFKZ,CardyBast,deter,KM,Smir2,DM,DSC} have used various
ways to identify and constrain the specific nature of the operator, e.g. by
looking at asymptotic behaviours, performing perturbation theory, taking
symmetries into account, formulating quantum equations of motion, etc. Our
analysis will make especially exploit the conjecture that each local
operator has a counterpart in the ultraviolet conformal field theory.

In the present manuscript we show for a concrete model, the SU(3)$_{2}$%
-homogeneous Sine-Gordon model (HSG), that, by means of the form factor
program, it is possible to reconstruct the field content starting from its
scattering matrix. Our analysis is based on the assumption \cite{KM,Smir2}
that each solution to the form factor consistency equations \cite
{Kar,Smir,Zamocorr,BFKZ} corresponds to a particular local operator. We take
furthermore into account that the SU(3)$_{2}$-HSG model, like numerous other
1+1 dimensional integrable models, may be viewed as a perturbed conformal
field theory whose entire field content is well classified. Assuming \ now a
one-to-one correspondence between operators in the conformal and in the
perturbed theory, we can carry out an identification on this level, that is
we associate each  solution of the form factor consistency equations a local
operator which is labeled according to the the ultra-violet conformal field
theory. We therefore construct systematically all possible solutions for the 
$n$-particle form factors related to a huge class of operators in terms of
some general building blocks which consist out of determinants of matrices
whose entries are elementary symmetric polynomials depending on the
rapidities. We demonstrate how these general solutions are interrelated by
the momentum space cluster property. In particular we show that the cluster
property serves also as a construction principle, in the sense that from one
solution to the consistency equations we may obtain a huge class, almost
all, of new solutions. Finally we compute the corresponding two-point
correlation functions and carry out the ultraviolet limit in order to
identify the corresponding conformal dimensions.

Our manuscript is organized as follows: In section 2 we recall \cite{CFK}
the solutions for the minimal form factors and the recursive equation which
is central for the determination of the form factors. We describe the
general structure of the $n$-particle form factors for a huge class of
operators. In section 3 we provide a rigorous proof for all solutions. In
section 4 we investigate the cluster property. In section 5 we compute
several two-point correlation functions and carry out the ultraviolet limit
on the base of a sum rule and the explicit two-point correlation function in
order to identify the conformal dimensions of each operator. We state our
conclusions in section 6. The appendix contains a collection of useful
properties of elementary symmetric polynomials and some explicit formulae
for the first non-vanishing form factors.

\section{The SU(3)$_{2}$-HSG model form factors}

The SU(3)$_{2}$-HSG model contains only two self-conjugate solitons which we
denote, following the conventions of \cite{CFK}, by ``$+$'' and ``$-$''. The
two particle scattering matrix as a function of the rapidity $\theta $
related to this model was found \cite{HSGS} to be 
\begin{equation}
S_{\pm \pm }=-1\quad \quad \text{and}\quad \quad S_{\pm \mp }(\theta )=\pm
\tanh \frac{1}{2}\left( \theta \pm \sigma -i\frac{\pi }{2}\right) \,\,.\;
\label{ZamS}
\end{equation}
Here $\sigma $ is a real constant and corresponds to a resonance parameter.
The system (\ref{ZamS}) constitutes probably the simplest example of a
massive quantum field theory involving two particles of distinct type.
Nonetheless, despite the simplicity of the scattering matrix we expect to
find a relatively involved operator content, since for finite resonance
parameter the SU(3)$_{2}$-HSG model describes a WZNW-coset model with
central charge $c=6/5$ perturbed by an operator with conformal dimension $%
\Delta =3/5$. Since this is true as long as $\sigma $ is finite, we shall be
content in the following mostly by setting $\sigma =0$. It is expected from
the classical analysis that for finite value of $\sigma $ we always have the
same ultraviolet central charge and therefore the same operator content. The
TBA-analysis carried out in \cite{CFKM} supports this analysis. Thus, a
finite variation of $\sigma $ at the ultraviolet fixed point and away from
it is not very illuminating and we therefore only distinguish the behaviour $%
\sigma \rightarrow \infty $ and $\sigma $ finite. In the former case one
trivially observes that the S-matrices S$_{\pm \mp }$ tend to one and the
theory decouples into two Ising models. The related form factors have to
respect this behaviour and all combinations involving different types of
particles vanish, see \cite{CFK}. One may see also directly that the form
factor solutions  behave this way by employing the Riemann-Lebesgue theorem. 

The underlying conformal field theory has recently \cite{Schou} found an
interesting application in the context of the construction of quantum Hall
states which carry a spin and fractional charges.

Taking the scattering matrix as an input, it is in principle possible to
compute form factors, by solving certain consistency equations \cite
{Kar,Smir,Zamocorr,BFKZ}, and thereafter to evaluate correlation functions.
Form factors are tensor valued functions, representing matrix elements of
some local operator $\mathcal{O}(\vec{x})$ located at the origin between a
multiparticle in-state and the vacuum, which we denote by

\begin{equation}
F_{n}^{\mathcal{O}|\mu _{1}\ldots \mu _{n}}(\theta _{1},\ldots ,\theta
_{n}):=\left\langle 0|\mathcal{O}(0)|V_{\mu _{1}}(\theta _{1})V_{\mu
_{2}}(\theta _{2})\ldots V_{\mu _{n}}(\theta _{n})\right\rangle _{\text{in}%
}\,\,.
\end{equation}
Here the $V_{\mu }(\theta )$ are some vertex operators representing a
particle of species $\mu $. We commence now by recalling the basic ansatz
for solutions of the form factors for the SU(3)$_{2}$-HSG model from \cite
{CFK}. We used the parameterization 
\begin{equation}
F_{n}^{\mathcal{O}|\stackrel{l\,\times \,\pm }{\overbrace{\mu _{1}\ldots \mu
_{l}}}\stackrel{m\,\times \,\mp }{\overbrace{\mu _{l+1}\ldots \mu _{n}}}%
}(\theta _{1}\ldots \theta _{n})=H_{n}^{\mathcal{O}|\mu _{1}\ldots \mu
_{n}}Q_{n}^{\mathcal{O}|\mu _{1}\ldots \mu _{n}}(x_{1}\ldots
x_{n})\,\!\!\prod_{i<j}\hat{F}^{\mu _{i}\mu _{j}}(\theta _{ij})  \label{fact}
\end{equation}
where $\hat{F}^{\mu _{i}\mu _{j}}(\theta _{ij}):=F_{\text{min}}^{\mu _{i}\mu
_{j}}(\theta _{ij})/\left( x_{i}^{\mu _{i}}+x_{j}^{\mu _{j}}\right) ^{\delta
_{\mu _{i}\mu _{j}}}$. The rapidities enter through the variable $x_{i}=\exp
(\theta _{i})$ and the functions $F_{\text{min}}^{\mu _{i}\mu _{j}}(\theta
_{ij})$ denote the so-called minimal form factors. They were found to be 
\begin{eqnarray}
F_{\text{min}}^{\pm \pm }(\theta ) &=&-i\sinh \frac{\theta }{2} \\
F_{\text{min}}^{\pm \mp }(\theta ) &=&2^{\frac{1}{4}}e^{\tfrac{i\pi (1\mp
1)\pm \theta }{4}-\tfrac{G}{\pi }}\exp \left( -\int\limits_{0}^{\infty }%
\tfrac{dt}{t}\tfrac{\sin ^{2}\left( (i\pi -\theta \mp \sigma )\frac{t}{2\pi }%
\right) }{\sinh t\cosh t/2}\right) \,=e^{\pm \tfrac{\theta }{4}}\tilde{F}_{%
\text{min}}^{\pm \mp }(\theta )\,\,.  \label{14}
\end{eqnarray}
Here $G$ is the Catalan constant. For the overall constants we obtained 
\begin{equation}
H^{\mathcal{O}|2s+\tau ,m}=i^{sm}2^{s(2s-m-1+2\tau )}e^{sm\sigma /2}H^{%
\mathcal{O}|\tau ,m},\qquad \tau =0,1\,.  \label{consol}
\end{equation}
Note that at this point an unknown constant, that is $H^{\mathcal{O}|\tau
,m} $, enters into the procedure. This quantity is not constrained by the
form factor consistency equations and has to be obtained from elsewhere. The
polynomials $Q$ have to satisfy the recursive equations

\begin{eqnarray}
Q^{\mathcal{O}|l+2,m}(-x,x,\ldots ,x_{n}) &=&D_{\vartheta
}^{l,m}(x,x_{1},\ldots ,x_{n})Q^{\mathcal{O}|l,m}(x_{1},\ldots ,x_{n})
\label{Qrec} \\
D_{\vartheta }^{l,m}(x,x_{1},\ldots ,x_{n}) &=&\frac{1}{2}(-ix)^{l+1}\sigma
_{l}^{+}\sum_{k=0}^{m}x^{-k}(1-(-1)^{l+k+\vartheta })\hat{\sigma}%
_{k}^{-}\,\,.
\end{eqnarray}
In particular 
\begin{equation}
D_{\zeta }^{2s+\tau ,2t+\tau ^{\prime }}(x,x_{1},\ldots
,x_{n})=(-i)^{2s+\tau +1}\sigma _{2s+\tau }^{+}\sum_{p=0}^{t}x^{2s-2p+\tau
+1-\zeta }\hat{\sigma}_{2p+\zeta }^{-}\,\,.  \label{Dpar}
\end{equation}
Here $\vartheta $ is related to the factor of local commutativity $\omega
=(-1)^{\vartheta }=\pm 1$. We introduced also the function $\zeta $ which is 
$0$ or $1$ for the sum $\vartheta +\tau $ being odd or even, respectively.
We shall use various notations for elementary symmetric polynomials (see
appendix for some essential properties). We employ the symbol $\sigma _{k}$
when the polynomials depend on the variables $x_{i}$, the symbol $\bar{\sigma%
}_{k}$ when they depend on the inverse variables $x_{i}^{-1}$, the symbol $%
\hat{\sigma}_{k}$ when they depend on the variables $x_{i}e^{-\sigma +i\pi
/2}$ and $\tilde{\sigma}_{k}$ when we set the first two variables to $%
x_{1}=-x,$ $x_{2}=x$. The number of variables the polynomials depend upon is
defined always in an unambiguous way through the l.h.s. of our equations,
where we assume the first $l$ variables to be associated with $\mu =+$ and
the last $m$ variables with $\mu =-$. In case no superscript is attached to
the symbol the polynomials depend on all $m+l$ variables, in case of a ``$+$%
'' they depend on the first $l$ variables and in case of a ``$-$'' on the
last $m$ variables.

Solving recursive equations of the type (\ref{Qrec}) in complete generality
is still an entirely open problem. Ideally one would like to reach a
situation similar to the one in the boostrap construction procedure of the
scattering matrices, where one can state general building blocks, e.g.
particular combinations of hyperbolic functions whenever backscattering is
absent \cite{Mitra}, infinite products of gamma functions when
backscattering occurs or elliptic functions when infinite resonances are
present. At least for all operators in the model we consider here this goal
has been achieved. It will turn out that all solutions to the recursive
equations (\ref{Qrec}) may be constructed from some general building blocks
consisting out of determinants of matrices whose entries are elementary
symmetric polynomials in some particular set of variables. Let us therefore
define the ($t+s$)$\times $($t+s$)-matrix 
\begin{equation}
\mathcal{\,}\left( \mathcal{A}_{l,m}^{\mu ,\nu }(s,t)\right)
_{ij}:=\QATOPD\{ . {\sigma _{2(j-i)+\mu }^{+}\qquad \,\,\,\,\,\,\,\,\text{%
for\quad }1\leq i\leq t}{\hat{\sigma}_{2(j-i)+2t+\nu }^{-}\,\,\,\quad \quad
\,\,\,\,\,\,\,\text{for\quad }t<i\leq s+t}\,\,.  \label{Acomp}
\end{equation}
More explicitly the matrix $\mathcal{A}$ reads 
\begin{equation}
\mathcal{A}_{l,m}^{\mu ,\nu }=\left( 
\begin{array}{rrrrrr}
\sigma _{\mu }^{+} & \sigma _{\mu +2}^{+} & \sigma _{\mu +4}^{+} & \sigma
_{\mu +6}^{+} & \cdots & 0 \\ 
0 & \sigma _{\mu }^{+} & \sigma _{\mu +2}^{+} & \sigma _{\mu +4}^{+} & \cdots
& 0 \\ 
\vdots & \vdots & \vdots & \vdots & \ddots & \vdots \\ 
0 & 0 & 0 & 0 & \cdots & \sigma _{2s+\mu }^{+} \\ 
\hat{\sigma}_{\nu }^{-} & \hat{\sigma}_{\nu +2}^{-} & \hat{\sigma}_{\nu
+4}^{-} & \hat{\sigma}_{\nu +6}^{-} & \cdots & 0 \\ 
0 & \hat{\sigma}_{\nu }^{-} & \hat{\sigma}_{\nu +2}^{-} & \hat{\sigma}_{\nu
+4}^{-} & \cdots & 0 \\ 
\vdots & \vdots & \vdots & \vdots & \ddots & \vdots \\ 
0 & 0 & 0 & 0 & \cdots & \hat{\sigma}_{2t+\nu }^{-}
\end{array}
\right) \,\,.  \label{sss}
\end{equation}

\noindent The superscripts $\mu ,\nu $ may take the values $0$ and $1$ and
the subscripts $l,m$ characterize the number of different variables related
to the particle species ``$+$'', ``$-$'', respectively. 
The different combinations of the integers $\mu ,\nu ,l,m$ will
correspond to different kind of local operators $\mathcal{O}$.

\noindent In addition,
the form factors will involve a function depending on two further indices $%
\bar{\mu}$ and $\bar{\nu}$%
\begin{equation}
g_{l,m}^{\bar{\mu},\bar{\nu}}:=(\sigma _{l})^{\frac{l-m+\bar{\mu}}{2}%
}(\sigma _{m})^{\frac{\bar{\nu}-m}{2}}\,\,.  \label{g}
\end{equation}
Here the $\bar{\mu},\bar{\nu}$ are integers whose range, unlike the one for $%
\mu ,\nu $, is in principle not restricted. However, it will turn out that
due to the existence of certain constraining relations, to be specified in
detail below, it is sufficient to characterize a particular operator by the
four integers $\mu ,\nu ,l,m$ only. Then, as we shall demonstrate, all $Q$%
-polynomials acquire the general form 
\begin{equation}
Q^{\mathcal{O}|l,m}=Q_{l,m}^{\mu ,\nu }=Q_{2s+\tau ,2t+\tau ^{\prime }}^{\mu
,\nu }=i^{s\nu }(-1)^{s(\tau +t+1)}g_{2s+\tau ,2t+\tau ^{\prime }}^{\bar{\mu}%
,\bar{\nu}}\,\,\det \mathcal{A}_{2s+\tau ,2t+\tau ^{\prime }}^{\mu ,\nu }\,.
\label{qpar}
\end{equation}

\noindent We used here already a parameterization for $l,m$ which will turn
out to be most convenient. The subscripts in $g$ and $\mathcal{A}$ are only
needed in formal considerations, but in most cases the number of particles
of species ``$+$'' and ``$-$'' are unambiguously defined through the l.h.s.
of our equations. This is in the same spirit in which we refer to the number
of variables in the elementary symmetric polynomials. We will therefore drop
them in these cases, which leads to simpler, but still precise, notations.
To illustrate this with examples, we consider for instance the solutions to
the recursive equations (\ref{Qrec}) related to the trace of the energy
momentum tensor $\Theta $, the order operator $\Sigma $ and the disorder
operator $\mu $ which were already stated in \cite{CFK} 
\begin{eqnarray}
Q^{\Theta |2s+2,2t+2} &=&i^{s(2t+3)}e^{-(t+1)\sigma }\sigma _{1}\bar{\sigma}%
_{1}\,g^{0,2}\,\det \mathcal{A}^{1,1}\mathcal{\,\,}  \label{solu1} \\
Q^{\Sigma |2s,2t+1} &=&i^{s(2t+3)}\,g^{-1,1}\,\det \mathcal{A}^{0,1}\mathcal{%
\,}  \label{solu2} \\
Q^{\mu |2s,2t} &=&i^{2s(t+1)}\,g^{-1,1}\,\det \mathcal{A}^{0,0}\mathcal{\,}.
\label{solu3}
\end{eqnarray}
Here $\mathcal{A}$ is always taken to be a ($t+s$)$\times $($t+s$)-matrix.
Notice that in comparison with (\ref{qpar}) the factor of proportionality in
(\ref{solu2}) and (\ref{solu3}) is only a constant, whereas in (\ref{solu1})
also the term $\sigma _{1}\bar{\sigma}_{1}$ appears. Terms of this type may
always be added since they satisfy the consistency equations trivially. This
is also the reason why in comparison with \cite{CFK} we can safely drop in $%
Q^{\Sigma |2s,2t+1}$ the factor $(\sigma _{1})^{\frac{1}{2}}(\sigma
_{1}^{-})^{-\frac{1}{2}}$. Additional reasons for this modification will be
provided below. For $\Theta $ we were forced \cite{CFK} to introduce the
factor $\sigma _{1}\bar{\sigma}_{1}$ in order to recover the solution of the
thermally perturbed Ising model for $2s+2=0$. Note that for $\Theta $ the
value $s=-1$ formally makes sense.

\section{Solution procedure}

We shall now recall the principle steps of the general solution procedure
for the form factor consistency equations \cite{Kar,Smir,Zamocorr,BFKZ}. For
any local operator $\mathcal{O}$ one may anticipate the pole structure of
the form factors and extract it explicitly in form of an ansatz of the type (%
\ref{fact}). This might turn out to be a relatively involved matter due to
the occurrence of higher order poles in some integrable theories, e.g. \cite
{DM}, but nonetheless it is possible. Thereafter the task of finding
solutions may be reduced to the evaluation of the minimal form factors and
to solving a (or two if bound states may be formed in the model) recursive
equation of the type (\ref{Qrec}). The first task can be carried out
relatively easily, especially if the related scattering matrix is given as a
particular integral representation \cite{Kar}. Then an integral
representation of the type (\ref{14}) can be deduced immediately. The second
task is rather more complicated and the heart of the whole problem. Having a
seed for the recursive equation, that is the lowest non-vanishing form factor%
\footnote{%
For the case at hand this is provided for some operators by the well known
solutions of the Ising model. In general this is also a difficult hurdle to
take as, for instance, one might need to know vacuum expectation values.},
one can in general compute from them several form factors which involve more
particles. However, the equations become relatively involved after several
steps. Aiming at the solution for all $n$-particle form factors, it is
therefore highly desirable to unravel a more generic structure which enables
one to formulate rigorous proofs. Several examples \cite
{clust,Zamocorr,deter} have shown that often the general solution may be
cast into the form of determinants whose entries are elementary symmetric
polynomials. Presuming such a structure which, at present, may be obtained
by extrapolating from lower particle solutions to higher ones or by some
inspired guess, one can rigorously formulate proofs as we now demonstrate
for the SU(3)$_{2}$-HSG-model, for which some solutions were merely stated
in \cite{CFK}.

We have two universal structures\footnote{%
There exist also different types of universal expressions like for instance
the integral representations presented in \cite{BFKZ}. However, these type
of expressions are sometimes only of a very formal nature since to evaluate
them concretely for higher $n$-particle form factors requires still a
considerable amount of computational effort.} at our disposal. We could
either exploit the integral representation for the determinant $\mathcal{A}$%
, as presented in \cite{CFK}, or exploit simple properties of determinants.
Here we shall pursue the latter possibility. For this purpose it is
convenient to define the operator $C_{i,j}^{x}$ $(R_{i,j}^{x})$ which acts
on the $j^{\text{th}}$ column (row) of an ($n\times n$)-matrix $\mathcal{A}$
by adding $x$ times the $i^{\text{th}}$ column (row) to it 
\begin{eqnarray}
C_{i,j}^{x}\mathcal{A} &:&\qquad \mathcal{A}_{kj}\mapsto \mathcal{A}_{kj}+x%
\mathcal{A}_{ki}\qquad \quad 1\leq i,j,k\leq n \\
R_{i,j}^{x}\mathcal{A} &:&\qquad \mathcal{A}_{jk}\mapsto \mathcal{A}_{jk}+x%
\mathcal{A}_{ik}\qquad \quad 1\leq i,j,k\leq n\,\,.
\end{eqnarray}
Naturally the determinant of $\mathcal{A}$ is left invariant under the
actions of $C_{i,j}^{x}$ and $R_{i,j}^{x}$ on $\mathcal{A},$ such that we
can use them to bring $\mathcal{A}$ into a suitable form for our purposes.
Furthermore, it is convenient to define the ordered products, i.e. operators
related to the lowest entry act first, 
\begin{equation}
\mathcal{C}_{a,b}^{x}:=\prod_{p=a}^{b}\!\!\,\,C_{p,p+1}^{x}\qquad \text{%
and\qquad }\mathcal{R}_{a,b}^{x}:=\prod_{p=a}^{b}\!\!\,\,R_{p,p-1}^{x}\,\,.
\end{equation}
It will be our strategy to use these operators in such a way that we produce
as many zeros as possible in one column or row of a matrix of interest to
us. In order to satisfy (\ref{Qrec}) we have to set now the first variables
in $\mathcal{A}$ to $x_{1}=-x,\,x_{2}=x$, which we denote as $\tilde{%
\mathcal{A}}$ thereafter and relate the matrices $\tilde{\mathcal{A}}%
_{l+2,m}^{\mu ,\nu }$ and $\mathcal{A}_{l,m}^{\mu ,\nu }$. Taking relation (%
\ref{symp}) for the elementary symmetric polynomials into account, we can
bring $\tilde{\mathcal{A}}_{l+2,m}^{\mu ,\nu }$ into the form 
\begin{equation}
\left( \!\!\mathcal{R}_{t+2,s+t+1}^{-x^{2}}\mathcal{C}_{1,s+t-1}^{x^{2}}%
\tilde{\mathcal{A}}_{l+2,m}^{\mu ,\nu }\right) _{ij}\!=\left\{ 
\begin{array}{l}
\sigma _{2(j-i)+\mu }^{+}\qquad \qquad \qquad \,\,\,\,\,\,\,\,\,\,\,\,1\leq
i\leq t \\ 
\hat{\sigma}_{2(j-i)+2t+\nu }^{-}\qquad \quad \quad \,\,\quad
\,\,\,\,\,\,\,\,\,\,t<i\leq s+t \\ 
\sum\limits_{p=1}^{j}x^{2(j-p)}\hat{\sigma}_{2(p-s-1)+\nu }^{-}\quad
\,\,\,\,\,\,\,\,\,\,\,\,i=s+t+1
\end{array}
\right. .
\end{equation}
It is now crucial to note that since the number of variables has been
reduced by two, several elementary polynomials may vanish. As a consequence,
for $2s+2+\mu >l$ and $2t+2+\nu >m$, the last column takes on the simple
form 
\begin{equation}
\left( \!\!\mathcal{R}_{t+2,s+t+1}^{-x^{2}}\mathcal{C}_{1,s+t-1}^{x^{2}}%
\tilde{\mathcal{A}}_{l+2,m}^{\mu ,\nu }\right) _{i(s+t+1)}\!=\left\{ \QATOP{%
0\qquad \,\,\,\,\,\,\,\,\,\,\,\,\,\,\,\,\,\,\,\,\,\,\,\,\,\,\,\,\,\,\,1\leq
i\leq s+t}{\sum\limits_{p=0}^{t}x^{2(t-p)}\hat{\sigma}_{2p+\nu }^{-}\quad
\,\,\,\,\,\,i=s+t+1\quad }\right. .\,  \label{19}
\end{equation}
Therefore, developing the determinant of $\tilde{\mathcal{A}}_{l+2,m}^{\mu
,\nu }$ with respect to the last column, we are able to relate the
determinants of $\tilde{\mathcal{A}}_{l+2,m}^{\mu ,\nu }$ and $\mathcal{A}%
_{l,m}^{\mu ,\nu }$ as 
\begin{equation}
\det \tilde{\mathcal{A}}_{l+2,m}^{\mu ,\nu }=\left(
\sum\limits_{p=0}^{t}x^{2(t-p)}\hat{\sigma}_{2p+\nu }^{-}\right) \,\det 
\mathcal{A}_{l,m}^{\mu ,\nu }\,\,\,\,\,\,\,\,.  \label{arec}
\end{equation}
We are left with the task to specify the behaviour of the function $g$ with
respect to the ``reduction'' of the first two variables 
\begin{equation}
\tilde{g}_{l+2,m}^{\bar{\mu},\bar{\nu}}=i^{l-m+\bar{\mu}+2}\,x^{l-m+\bar{\mu}%
+2}\,\sigma _{l}^{+}\,g_{l,m}^{\bar{\mu},\bar{\nu}}\,\,\,.  \label{grec}
\end{equation}
Assembling the two factors (\ref{arec}) and (\ref{grec}), we obtain, in
terms of the parameterization (\ref{qpar}) 
\begin{equation}
\tilde{Q}_{2s+2+\tau ,2t+\tau ^{\prime }}^{\bar{\mu},\bar{\nu},\mu ,\nu
}=(-i)^{2s+\tau +1}\sigma _{2s+\tau }^{+}\left(
\sum\limits_{p=0}^{t}x^{2(s-p+1)+\tau -\tau ^{\prime }+\bar{\mu}}\hat{\sigma}%
_{2p+\nu }^{-}\right) \,Q_{2s+\tau ,2t+\tau ^{\prime }}^{\bar{\mu},\bar{\nu}%
,\mu ,\nu }\,\,.  \label{soll}
\end{equation}
We are now in the position to compare our general construction (\ref{soll})
with the recursive equation for the $Q$-polynomials of the SU(3)$_{2}$-HSG
model (\ref{Dpar}). We read off directly the following restrictions 
\begin{equation}
\nu =\zeta \qquad \text{and\qquad }\zeta =\tau ^{\prime }-\bar{\mu}-1\,\,\,.
\label{rest}
\end{equation}
A further constraint results from relativistic invariance, which implies
that the overall power in all variables $x_{i}$ of the form factors has to
be zero for a spinless operator. Introducing the short hand notation [$%
F_{n}^{\mathcal{O}}$] for the total power, we have to evaluate 
\begin{equation}
\left[ Q_{2s+\tau ,2t+\tau ^{\prime }}^{\bar{\mu},\bar{\nu},\mu ,\nu }\right]
=\,\left[ g_{2s+\tau ,2t+\tau ^{\prime }}^{\bar{\mu},\bar{\nu}}\right] +\,%
\left[ \det \mathcal{A}_{2s+\tau ,2t+\tau ^{\prime }}^{\mu ,\nu }\right]
\,\,.  \label{QT}
\end{equation}
Combining (\ref{rest}) and (\ref{QT}) with the explicit expressions $\,[\det 
\mathcal{A}_{2s+\tau ,2t+\tau ^{\prime }}^{\mu ,\nu }]=s(2t+\nu )+\mu t$, $%
\,[g_{l,m}^{\bar{\mu},\bar{\nu}}]=l(l-m+\bar{\mu})/2+m(\bar{\nu}-m)/2$ and $%
[Q_{2s+\tau ,2t+\tau ^{\prime }}^{\bar{\mu},\bar{\nu},\mu ,\nu
}]=l(l-1)/2-m(m-1)/2$, we find the additional constraints 
\begin{equation}
\mu =1+\tau -\bar{\nu}\qquad \text{and\qquad }\tau \nu =\tau ^{\prime }(\bar{%
\nu}-1)\,\,.  \label{rest2}
\end{equation}
Collecting now everything we conclude that different solutions to the form
factor consistency equations can be characterized by a set of four distinct
integers. Assuming that each solution corresponds to a local operator, there
might be degeneracies of course, we can label the operators by $\mu ,\nu
,\tau ,\tau ^{\prime }$, i.e. $\mathcal{O}\rightarrow \mathcal{O}_{\tau
,\tau ^{\prime }}^{\mu ,\nu }$, such that we can also write $Q_{m,l}^{\mu
,\nu }$ instead of $Q_{m,l}^{\bar{\mu},\bar{\nu},\mu ,\nu }$. Then each $Q$%
-polynomial takes on the general form 
\begin{equation}
Q^{\mathcal{O}|2s+\tau ,2t+\tau ^{\prime }}=Q^{\mathcal{O}_{\tau ,\tau
^{\prime }}^{\mu ,\nu }|2s+\tau ,2t+\tau ^{\prime }}=Q_{2s+\tau ,2t+\tau
^{\prime }}^{\mu ,\nu }\sim g_{2s+\tau ,2t+\tau ^{\prime }}^{\tau ^{\prime
}-1-\nu ,\tau +1-\mu }\,\,\det \mathcal{A}_{2s+\tau ,2t+\tau ^{\prime
}}^{\mu ,\nu }\,\,
\end{equation}
and the integers $\mu ,\nu ,\tau ,\tau ^{\prime }$ are restricted by 
\begin{equation}
\tau \nu +\tau ^{\prime }\mu =\tau \tau ^{\prime },\quad \qquad 2+\mu >\tau
,\quad \qquad 2+\nu >\tau ^{\prime }\,\,.  \label{25}
\end{equation}
We combined here (\ref{rest}) and (\ref{rest2}) to get the first relation in
(\ref{25}). The inequalities result from the requirement in the proof which
we needed to have the form (\ref{19}). We find 12 admissible solutions to (%
\ref{25}), i.e. potentially 12 different local operators, whose quantum
numbers are presented in table 1.

\begin{center}
\vspace{0.2cm} 
\begin{tabular}{|r|r|r|r|r|r|r|}
\hline
$\mu $ & $\nu $ & $\tau $ & $\tau ^{\prime }$ & $\left[ F_{\tau \tau
^{\prime }}^{\mu \nu }\right] _{+}$ & $\left[ F_{\tau \tau ^{\prime }}^{\mu
\nu }\right] _{-}$ & $\Delta $ \\ \hline\hline
0 & 0 & 0 & 0 & 0 & 0 & 1/10 \\ \hline
0 & 0 & 1 & 0 & 0 & 0 & 1/10 \\ \hline
0 & 0 & 0 & 1 & 0 & 0 & 1/10 \\ \hline
0 & 1 & 0 & 1 & -1/2 & 0 & 1/10 \\ \hline
0 & 1 & 1 & 1 & -1/2 & 0 & 1/10 \\ \hline
0 & 1 & 0 & 2 & -1/2 & 0 & 1/10 \\ \hline
1 & 0 & 1 & 1 & 0 & -1/2 & 1/10 \\ \hline
1 & 0 & 2 & 0 & 0 & -1/2 & 1/10 \\ \hline
1 & 0 & 1 & 0 & 0 & -1/2 & 1/10 \\ \hline
1 & 1 & 2 & 2 & -1/2 & -1/2 & 1/10 \\ \hline
1 & 1 & 0 & 0 & -1/2 & -1 & * \\ \hline
1 & 0 & 0 & 0 & 0 & -1/2 & * \\ \hline
\end{tabular}
\vspace{0.4cm}

{\small Table 1: Operator content of the $SU(3)_{2}$-HSG model.}
\end{center}

\noindent Comparing with our previous results, we have according to this
notation $F^{\mathcal{O}_{0,0}^{0,0}|2s,2t}=F^{\mu |2s,2t}$, $F^{\mathcal{O}%
_{0,1}^{0,1}|2s,2t+1}=F^{\Sigma |2s,2t+1}$ and $F^{\mathcal{O}%
_{2,2}^{1,1}|2s,2t+1}\sim F^{\Theta |2s+2,2t+2}$. The last two solutions are
only formal in the sense that they solve the constraining equations (\ref{25}%
), but the corresponding explicit expressions turn out to be zero.

In summary, by taking the determinant of the matrix (\ref{sss}) as the
ansatz for the general building block of the form factors, we constructed
systematically generic formulae for the $n$-particle form factors possibly
related to 12 different operators.

\section{Momentum space cluster properties}

Cluster properties in space, i.e. the observation that far separated
operators do not interact, are quite familiar in quantum field theories \cite
{Wich} for a long time. In 1+1 dimensions a similar property has also been
noted in momentum space. For the purely bosonic case this behaviour can be
explained perturbatively by means of Weinberg's power counting theorem, see
e.g. \cite{Kar,BK}\footnote{%
There exists also a heuristic argument which provides some form of intuitive
picture of this behaviour\cite{DSC} by appealing to the ultraviolet
conformal field theory. However, the argument is based on various assumptions,
which need further clarification. For instance it remains to be proven
rigorously that the particle creation operator $V_{\mu }(\theta )$ tends to
a conformal Zamolodchikov operator for $\theta \rightarrow \infty $ and that
the local field factorizes equally into two chiral fields in that situation.
The restriction in there that $\lim_{\theta \rightarrow \infty
}S_{ij}(\theta )=1$, for $i$ being a particle which has been shifted and $j$
one which has not, excludes a huge class of interesting models, in
particular the one at hand.}. This property has been analysed explicitly for
several specific models \cite{clust,Zamocorr,KM,MS}. It states that whenever
the first, say $\kappa $, rapidities of an $n$-particle form factor are
shifted to infinity, the $n$-particle form factor factorizes into a $\kappa $
and an ($n-\kappa $)-particle form factor which are possibly related to
different types of operators 
\begin{equation}
\mathcal{T}_{1,\kappa }^{\lambda }F_{n}^{\mathcal{O}}(\theta _{1},\ldots
,\theta _{n})\sim F_{\kappa }^{\mathcal{O}^{\prime }}(\theta _{1},\ldots
,\theta _{\kappa })F_{n-\kappa }^{\mathcal{O}^{\prime \prime }}(\theta
_{\kappa +1},\ldots ,\theta _{n})\,\,.  \label{cluster}
\end{equation}
For convenience we have introduced here the operator 
\begin{equation}
\mathcal{T}_{a,b}^{\lambda }:=\lim_{\lambda \rightarrow \infty
}\prod_{p=a}^{b}T_{p}^{\lambda }\,
\end{equation}
which will allow for concise notations. It is composed of the translation
operator $T_{a}^{\lambda }$ which acts on a function of $n$ variables as 
\begin{equation}
T_{a}^{\lambda }\,f(\theta _{1},\ldots ,\theta _{a},\ldots ,\theta
_{n})\,\mapsto \,f(\theta _{1},\ldots ,\theta _{a}+\lambda ,\ldots ,\theta
_{n})\,\,\,.
\end{equation}
Whilst Watson's equations and the residue equations, see e.g. \cite
{Kar,Smir,Zamocorr,CFK}, are operator independent features of form factors,
the cluster property captures part of the operator nature of the theory. The
cluster property (\ref{cluster}) does not only constrain the solution, but
eventually also serves as a construction principle in the sense that when
given $F_{n}^{\mathcal{O}}$ we may employ (\ref{cluster}) and construct form
factors related to $\mathcal{O}^{\prime }$ and $\mathcal{O}^{\prime \prime }$%
. Hence, (\ref{cluster}) constitutes a closed mathematical structure, which
relates various solutions and whose abstract nature still needs to be
unraveled.

We shall now systematically investigate the cluster property (\ref{cluster})
for the $SU(3)_{2}$-HSG model. Choosing w.l.g. the upper signs for the
particle types in equation (\ref{fact}), we have four different options to
shift the rapidities 
\begin{eqnarray}
\mathcal{T}_{1,\kappa \leq l}^{\pm \lambda }F_{n}^{\mathcal{O}|l\times
+,m\times -} &=&\mathcal{T}_{\kappa +1<l,n}^{\mp \lambda }F_{n}^{\mathcal{O}%
|l\times +,m\times -}  \label{s1} \\
\mathcal{T}_{1,\kappa >l}^{\pm \lambda }F_{n}^{\mathcal{O}|l\times +,m\times
-} &=&\mathcal{T}_{\kappa +1\geq l,n}^{\mp \lambda }F_{n}^{\mathcal{O}%
|l\times +,m\times -}\,  \label{s2}
\end{eqnarray}
which a priori might all lead to different factorizations on the r.h.s. of
equation (\ref{cluster}). The equality signs in the equations (\ref{s1}) and
(\ref{s2}) are a simple consequence of the relativistic invariance of form
factors, i.e. we may shift all rapidities by the same amount, for $\mathcal{O%
}$ being a scalar operator.

Considering now the ansatz (\ref{fact}) we may first carry out part of the
analysis for the terms which are operator independent. Noting that 
\begin{equation}
\mathcal{T}_{1,1}^{\pm \lambda }F_{\text{min}}^{++}(\theta )=\mathcal{T}%
_{1,1}^{\pm \lambda }F_{\text{min}}^{--}(\theta )\sim e^{\frac{(\lambda \pm
\theta )}{2}}\qquad \text{and\qquad }\mathcal{T}_{1,1}^{\pm \lambda }F_{%
\text{min}}^{+-}(\theta )\sim \QATOPD\{ . {\mathcal{O}(1)}{e^{\frac{(\theta
-\lambda )}{2}}}\,,  \label{eq}
\end{equation}
we obtain for the choice of the upper signs for the particle types in the
ansatz (\ref{fact}) 
\[
\mathcal{T}_{1,\kappa \leq l}^{\pm \lambda }\,\prod_{i<j}\hat{F}^{\mu
_{i}\mu _{j}}(\theta _{ij})\sim \!\!\!\!\!\!\prod_{1\leq i<j\leq \kappa
}\!\!\!\!\!\hat{F}^{++}(\theta _{ij})\!\!\!\!\!\!\!\prod_{\kappa <i<j\leq
l+m}\!\!\!\!\!\!\hat{F}^{\mu _{i}\mu _{j}}(\theta _{ij})\left\{ \QTATOP{%
\tfrac{\sigma _{\kappa }(x_{1},\ldots ,x_{\kappa })^{\frac{\kappa -l}{2}}e^{%
\frac{\lambda \kappa (1-l)}{2}}}{\sigma _{l-\kappa }(x_{\kappa +1},\ldots
,x_{l})^{\frac{\kappa }{2}}}}{\tfrac{\sigma _{\kappa }(x_{1},\ldots
,x_{\kappa })^{\frac{m-l+\kappa }{2}}e^{\frac{\lambda \kappa (l-m-1)}{2}}}{%
\sigma _{n-\kappa }(x_{\kappa +1},\ldots ,x_{n})^{\frac{\kappa }{2}}}}%
\right. 
\]
\[
\mathcal{T}_{n+1-\kappa <m,n}^{\pm \lambda }\,\prod_{i<j}\hat{F}^{\mu
_{i}\mu _{j}}(\theta _{ij})\sim \!\!\!\!\!\!\!\!\prod_{1\leq i<j\leq
n-\kappa }\!\!\!\!\!\!\!\hat{F}^{\mu _{i}\mu _{j}}(\theta
_{ij})\!\!\!\!\!\!\!\prod_{n-\kappa <i<j\leq n}\!\!\!\!\!\!\!\hat{F}%
^{--}(\theta _{ij})\left\{ \QATOP{\tfrac{\sigma _{n-\kappa }(x_{1},\ldots
,x_{n-\kappa })^{\frac{\kappa }{2}}e^{\frac{\lambda \kappa (m-l-1)}{2}}}{%
\sigma _{\kappa }(x_{n+1-\kappa },\ldots ,x_{n})^{\frac{l-m+\kappa }{2}}}}{%
\tfrac{\sigma _{m-\kappa }(x_{l+1},\ldots ,x_{n-\kappa })^{\frac{\kappa }{2}%
}e^{\frac{\lambda \kappa (1-m)}{2}}}{\sigma _{\kappa }(x_{n+1-\kappa
},\ldots ,x_{n})^{\frac{\kappa -m}{2}}}}\right. . 
\]
The remaining cases can be obtained from the equalities (\ref{s1}) and (\ref
{s2}). Turning now to the behaviour of the function $g$ as defined in (\ref
{g}) under these operations, we observe with help of the asymptotic
behaviour of the elementary symmetric polynomials (\ref{as+}) and (\ref{as-}%
) 
\begin{eqnarray}
\mathcal{T}_{1,\kappa \leq l}^{\pm \lambda }\,g_{l,m}^{\bar{\mu},\bar{\nu}%
}\!\!\! &=&\!\!\![e^{\pm \lambda \kappa }\sigma _{\kappa }(x_{1},\ldots
,x_{\kappa })\sigma _{l-\kappa }(x_{\kappa +1},\ldots ,x_{l})]^{\frac{l-m+%
\bar{\mu}}{2}}\,(\sigma _{m})^{\frac{\bar{\nu}-m}{2}} \\
\mathcal{T}_{n+1-\kappa <m,n}^{\pm \lambda }g_{l,m}^{\bar{\mu},\bar{\nu}%
}\!\!\! &=&\!\!\!(\sigma _{l})^{\frac{l-m+\bar{\mu}}{2}}[e^{\pm \lambda
\kappa }\sigma _{\kappa }(x_{n+1-\kappa },\ldots ,x_{n})\sigma _{m-\kappa
}(x_{l+1},\ldots ,x_{n-\kappa })]^{\frac{\bar{\nu}-m}{2}}.\,\,\,\,\,\,\,
\end{eqnarray}

\noindent In a similar fashion we compute the behaviour of the determinants 
\begin{eqnarray}
\mathcal{T}_{1,2\kappa +\xi \leq l}^{\lambda }\det \mathcal{A}_{l,m}^{\mu
,\nu } &=&e^{\lambda t(2\kappa +\xi )}(\sigma _{2\kappa +\xi })^{t}((-1)^{t}%
\hat{\sigma}_{\nu }^{-})^{\kappa +\xi (1-\mu )}\det \mathcal{A}_{l-2\kappa
-\xi ,m}^{1-\mu ,\nu } \\
\mathcal{T}_{1,2\kappa +\xi \leq l}^{-\lambda }\det \mathcal{A}_{l,m}^{\mu
,\nu } &=&(\hat{\sigma}_{2t+\nu }^{-})^{\kappa +\xi }\det \mathcal{A}%
_{l-2\kappa -\xi ,m}^{\mu ,\nu } \\
\mathcal{T}_{n+1-2\kappa -\xi <m,n}^{\lambda }\det \mathcal{A}_{l,m}^{\mu
,\nu } &=&e^{\lambda s(2\kappa +\xi )}(\sigma _{\mu }^{+})^{\kappa +\xi
(1-\nu )}(\hat{\sigma}_{2\kappa +\xi })^{s}\det \mathcal{A}_{l,m-2\kappa
-\xi }^{\mu ,1-\nu } \\
\mathcal{T}_{n+1-2\kappa -\xi <m,n}^{-\lambda }\det \mathcal{A}_{l,m}^{\mu
,\nu } &=&((-1)^{s}\sigma _{2s+\mu }^{+})^{\kappa +\xi }\det \mathcal{A}%
_{l,m-2\kappa -\xi }^{\mu ,\nu }\,\,.
\end{eqnarray}
We have to distinguish here between the odd and even case, which is the
reason for the introduction of the integer $\xi $ taking on the values $0$
or $1$. 

\vspace*{0.5cm}

\begin{center}
\includegraphics[width=14cm,height=14cm]{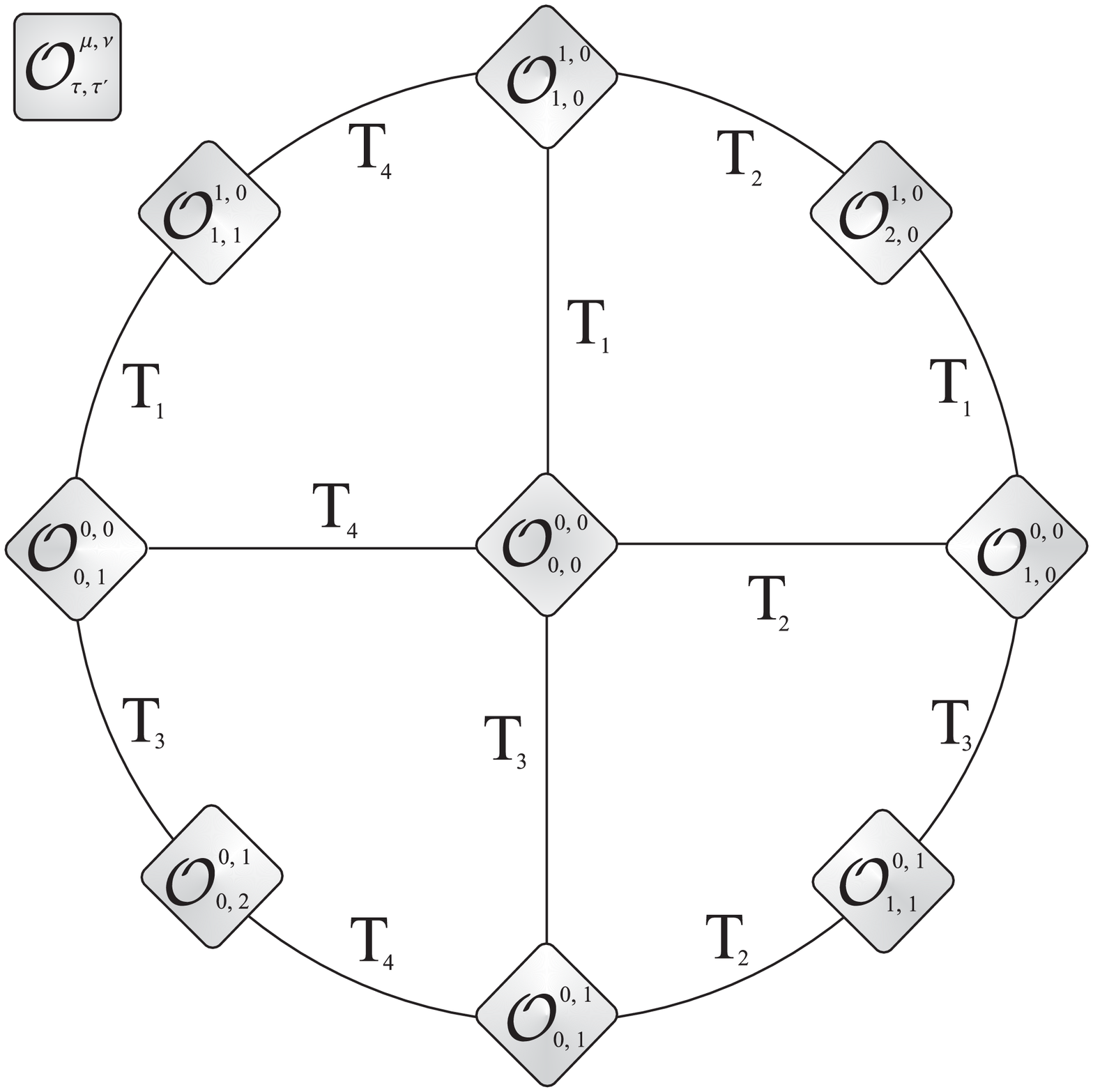}
\end{center}

\noindent {\small Figure 1: Interrelation of various operators via
clustering. In this figure we use the abbreviations $T_{1}\equiv \mathcal{T}%
_{1,2\kappa +1\leq l}^{\lambda }$, $T_{2}\equiv \mathcal{T}_{1,2\kappa
+1\leq l}^{-\lambda }$, $T_{3}\equiv \mathcal{T}_{n-2\kappa <m,n}^{\lambda }$%
, $T_{4}\equiv \mathcal{T}_{n-2\kappa <m,n}^{-\lambda }$. We also drop the
2s and 2t in the subscripts of the }$\mathcal{O}${\small 's. The }$T_{i}$ 
{\small on the links operate in both directions. } \vspace*{1.2mm}

Collecting now all the factors, we extract first the leading order
behaviour in $\lambda $%
\begin{equation}
\mathcal{T}_{1,\kappa \leq l}^{\pm \lambda }F_{2s+\tau ,2t+\tau ^{\prime
}}^{\mu ,\nu }\sim e^{-\lambda \kappa \left( \pm \nu +\tau ^{\prime }\frac{%
(1\mp 1)}{2}\right) }\qquad \mathcal{T}_{n+1-\kappa <m,n}^{\pm \lambda
}F_{2s+\tau ,2t+\tau ^{\prime }}^{\mu ,\nu }\sim e^{-\lambda \kappa \left(
\pm \mu +\tau \frac{(1\mp 1)}{2}\right) }\,.  \label{lead}
\end{equation}
Notice that, if we require that all possible actions of $\mathcal{T}%
_{a,b}^{\pm \lambda }$ should lead to finite expressions on the r.h.s. of (%
\ref{cluster}), we have to impose two further restrictions, namely $\mathcal{%
\tau }^{\prime }\geq \nu $ and $\tau \geq \mu $. These restrictions would
also exclude the last two solutions from table 1. We observe further that $%
F_{2,2}^{1,1}$ tends to zero under all possible shifts. Seeking now
solutions for the set $\mu ,\nu ,\tau ,\tau ^{\prime }$ of (\ref{lead})
which at least under some operations leads to finite results and in all
remaining cases tends to zero, we end up precisely with the first 9
solutions in table 1.

Concentrating now in more detail on these latter cases which behave like $%
\mathcal{O}(1)$, we find from the previous equations the following cluster
properties 
\begin{eqnarray}
\mathcal{T}_{1,2\kappa +\xi \leq l}^{\lambda }F_{2s+\tau ,2t+\tau ^{\prime
}}^{\mu ,0} &\sim &F_{2\kappa +\xi ,0}^{0,0}F_{2s+\tau -2\kappa -\xi
,2t+\tau ^{\prime }}^{\mu +\xi (1-2\mu ),0}  \label{40} \\
\mathcal{T}_{1,2\kappa +\xi \leq l}^{-\lambda }F_{2s+\tau ,2t+\nu }^{\mu
,\nu } &\sim &F_{2\kappa +\xi ,0}^{0,0}F_{2s+\tau -2\kappa -\xi ,2t+\nu
}^{\mu ,\nu } \\
\mathcal{T}_{n+1-2\kappa -\xi <m,n}^{\lambda }F_{2s+\tau ,2t+\tau ^{\prime
}}^{0,\nu } &\sim &F_{2s+\tau ,2t+\tau ^{\prime }-2\kappa -\xi }^{0,\nu +\xi
(1-2\nu )}F_{0,2\kappa +\xi }^{0,0} \\
\mathcal{T}_{n+1-2\kappa -\xi <m,n}^{-\lambda }F_{2s+\mu ,2t+\tau ^{\prime
}}^{\mu ,\nu } &\sim &F_{2s+\mu ,2t+\tau ^{\prime }-2\kappa -\xi }^{\mu ,\nu
}F_{0,2\kappa +\xi }^{0,0}\,\,.  \label{41}
\end{eqnarray}
We may now use (\ref{40})-(\ref{41}) as a means of constructing new
solutions, i.e. we can start with one solution and use (\ref{40})-(\ref{41})
in order to obtain new ones. Figure 1 demonstrates that when knowing just
one of the first nine operators in table 1 it is possible to (re)-construct
all the others in this fashion.

\subsection{The energy momentum tensor}

As we observed from our previous discussion the solution $F_{2,2}^{1,1}$ is
rather special. In fact this solution is part of the expression which in 
\cite{CFK} was identified as the trace of the energy momentum tensor 
\begin{equation}
Q^{\Theta |2s+2,2t+2}=i^{s(2t+3)}e^{-(t+1)\sigma }\sigma _{1}\bar{\sigma}%
_{1}F_{2s+2,2t+2}^{1,1}\,\,.
\end{equation}
The pre-factor $\sigma _{1}\bar{\sigma}_{1}$ will, however, alter the
cluster property. The leading order behaviour reads now 
\begin{equation}
\mathcal{T}_{1,\kappa \leq 2s}^{\pm \lambda }F^{\Theta |2s,2t}\sim \mathcal{T%
}_{n+1-\kappa <2t,n}^{\pm \lambda }F^{\Theta |2s,2t}\sim e^{\lambda
(1-\kappa /2)}\,\,.
\end{equation}
We observe that still in most cases the shifted expressions tend to zero,
unless $\kappa =1$ for which it tends to infinity as a consequence of the
introduction of the $\sigma _{1}\bar{\sigma}_{1}$. There is now also the
interesting case $\kappa =2$, for which the $\lambda $-dependence drops out
completely. Considering this case in more detail we find 
\begin{eqnarray}
\mathcal{T}_{1,2}^{\pm \lambda }F^{\Theta |2s,2t} &\sim &F^{\Theta
|2,0}F^{\Theta |2s-2,2t}\times \QATOPD\{ . {\frac{\sigma
_{1}(x_{2s+1},\ldots ,x_{2s+2t})}{\sigma _{1}(x_{3},\ldots ,x_{2s+2t})}}{%
\frac{\bar{\sigma}_{1}(x_{2s+1},\ldots ,x_{2s+2t})}{\bar{\sigma}%
_{1}(x_{3},\ldots ,x_{2s+2t})}}  \label{fa1} \\
\mathcal{T}_{n+1-\kappa ,n}^{\pm \lambda }F^{\Theta |2s,2t} &\sim &F^{\Theta
|2s,2t-2}F^{\Theta |0,2}\times \QATOPD\{ . {\frac{\sigma _{1}(x_{1},\ldots
,x_{2s})}{\sigma _{1}(x_{1},\ldots ,x_{2s+2t-2})}}{\frac{\bar{\sigma}%
_{1}(x_{1},\ldots ,x_{2s})}{\bar{\sigma}_{1}(x_{1},\ldots ,x_{2s+2t-2})}}\,.
\label{fa2}
\end{eqnarray}

\noindent Note that unless $s=1$ in (\ref{fa1}) or $t=1$ in (\ref{fa2}) the
form factors do not ``purely'' factorize into known form factors, but in all
cases a parity breaking factor emerges.

\noindent We now turn to the cases $\kappa =2s$ or $\kappa =2t$ for which we
derive 
\begin{equation}
\mathcal{T}_{1,2s}^{\pm \lambda }F^{\Theta |2s,2t}\sim \mathcal{T}%
_{n+1-2t,n}^{\pm \lambda }F^{\Theta |2s,2t}\sim e^{\lambda (2-t-s)}\,\,.
\end{equation}
We observe that once again in most cases these expressions tend to zero.
However, we also encounter several situations in which the $\lambda $%
-dependence drops out altogether. It may happen whenever $t=2$, $s=0$ or $%
s=2 $, $t=0$, which simply expresses the relativistic invariance of the form
factor. The other interesting situation occurs for $t=1$, $s=1$. Choosing
temporarily (in general we assume $m_{-}=m_{+}$) $H_{2}^{\Theta |0,2}=2\pi
m_{-}^{2}$ , $m=m_{-}=m_{+}e^{2G/\pi }$, we derive in this case 
\begin{equation}
\mathcal{T}_{1,2}^{\lambda }F_{4}^{\Theta |2,2}=\frac{F_{2}^{\Theta
|2,0}F_{2}^{\Theta |0,2}}{2\pi m^{2}}\,\,.
\end{equation}
In general when shifting the first $2s$ or last $2t$ rapidities we find the
following factorization 
\begin{equation}
\mathcal{T}_{1,2s}^{\pm \lambda }F^{\Theta |2s,2t}\sim \mathcal{T}%
_{n+1-2t,n}^{\pm \lambda }F^{\Theta |2s,2t}\sim \text{ }F^{\Theta
|2s,0}F^{\Theta |0,2t}\,\,.
\end{equation}
This equation holds true when keeping in mind that the r.h.s. of this
equation vanishes once it involves a form factor with more than two
particles. Note that only in these two cases the form factors factorize
``purely'' into two form factors without the additional parity breaking
factors as in (\ref{fa1}) and (\ref{fa2}).

\section{Identifying the operator content}

Having solved Watson's and the residue equations one has still little
information about the precise nature of the operator corresponding to a
particular solution. There exist, however, various non-perturbative (in the
standard coupling constant sense) arguments which provide this additional
information and which we now wish to exploit for the model at hand.
Basically all these arguments rely on the assumption that the superselection
sectors of the underlying conformal field theory remain separated after a
mass scale has been introduced. We will therefore first have a brief look at
the operator content of the $G_{k}/U(1)^{r}$-WZNW coset models and attempt
thereafter to match them with the solutions of the form factor consistency
equations. For these theories the different conformal dimensions in one
model can be parameterized by two quantities \cite{Confdim}: a highest
dominant weight $\Lambda $ of level smaller or equal to $k$ and their
corresponding lower weights $\lambda $ obtained in the usual way by
subtracting multiples of simple roots $\alpha _{i}$ from $\Lambda $ until
the lowest weight is reached 
\begin{equation}
\Delta (\Lambda ,\lambda )=\frac{(\Lambda \cdot (\Lambda +2\rho ))}{2(k+h)}-%
\frac{(\lambda \cdot \lambda )}{2k}\,\,.  \label{wei}
\end{equation}
Here $h$ is the Coxeter number of $G$ and $\rho $ the Weyl vector, i.e. the
sum over all fundamental weights. Denoting the highest root of $G$ by $\psi $%
, the conformal dimension related to the adjoint representation $\Delta
(\psi ,0)$ is of special interest since it corresponds to the one of the
perturbing operator which leads to the massive HSG-models. Taking the length
of $\psi $ to be $2$ and recalling the well known fact that the height of $%
\psi $, that is ht($\psi $), is the Coxeter number minus one, such that $%
(\psi \cdot \rho )=ht(\psi )=h-1$, it follows that $O^{\Delta (\psi ,0)}$ is
a unique operator with conformal dimension $\Delta (\psi ,0)=h/(k+h)$.%
{\small \ }Note that uniqueness demands in addition that we do not take the
multiplicities of the $\lambda $-states into account. For $SU(3)_{2}$ the
expression (\ref{wei}) is easily computed and since we could not find the
explicit values in the literature we report them for reference in table 2.

\begin{center}
\begin{tabular}{|c||c|c|c|c|c|}
\hline
$\lambda \backslash \Lambda $ & $\lambda _{1}$ & $\lambda _{2}$ & $\lambda
_{1}+\lambda _{2}$ & $2\lambda _{1}$ & $2\lambda _{2}$ \\ \hline\hline
$\Lambda $ & $1/10$ & $1/10$ & $1/10$ & $0$ & $0$ \\ \hline
$\Lambda -\alpha _{1}$ & $1/10$ & $*$ & $1/10$ & $1/2$ & $*$ \\ \hline
$\Lambda -\alpha _{2}$ & $*$ & $1/10$ & $1/10$ & $*$ & $1/2$ \\ \hline
$\Lambda -\alpha _{1}-\alpha _{2}$ & $1/10$ & $1/10$ & $3/5$ & $1/2$ & $1/2$
\\ \hline
$\Lambda -2\alpha _{1}$ & $*$ & $*$ & $*$ & $0$ & $*$ \\ \hline
$\Lambda -2\alpha _{2}$ & $*$ & $*$ & $*$ & $*$ & $0$ \\ \hline
$\Lambda -2\alpha _{1}-\alpha _{2}$ & $*$ & $*$ & $1/10$ & $1/2$ & $*$ \\ 
\hline
$\Lambda -\alpha _{1}-2\alpha _{2}$ & $*$ & $*$ & $1/10$ & $*$ & $1/2$ \\ 
\hline
$\Lambda -2\alpha _{1}-2\alpha _{2}$ & $*$ & $*$ & $1/10$ & $0$ & $0$ \\ 
\hline
\end{tabular}
\vspace{0.4cm}

{\small Table 2: Conformal dimensions for $\mathcal{O}^{\Delta (\Lambda
,\lambda )}$ in the $SU(3)_{2}/U(1)^{2}$ -coset model.}
\end{center}

\noindent Turning now to the massive theory, a crude constraint which gives
a first glimpse at possible solutions to the form factor consistency
equations is provided by the bound \cite{DM} 
\begin{equation}
\left[ F_{n}^{\mathcal{O}|\mu _{1}\ldots \mu _{n}}(\theta _{1},\ldots
,\theta _{n})\right] _{i}\leq \Delta ^{\!\!\mathcal{O}}\,\,\,.  \label{bound}
\end{equation}
We introduced here $\lim_{\theta _{i}\rightarrow \infty }f(\theta
_{1},\ldots ,\theta _{n})=:$ const $\exp ([f(\theta _{1},\ldots ,\theta
_{n})]_{i}\theta _{i})$ as abbreviation and denote the conformal dimension
of the operator $\mathcal{O}$ in the ultraviolet conformal limit by $\Delta
^{\!\!\mathcal{O}}$. We use the notation $[\,\,\,]_{\pm }$ when we take the
limit in the variable $x_{i}$ related to the particle species $\mu
_{i}=``\pm "$, respectively. For the different solutions we constructed, we
report the asymptotic behaviour in table 1. When we are in a position in
which we already anticipate the conformal dimensions the bound (\ref{bound})
will severely restrict the possible inclusion of factors like $\sigma _{1},%
\bar{\sigma}_{1},\sigma _{1}^{-},\sigma _{1}^{+}$, which as we mentioned
above may always be added since they trivially satisfy the consistency
equations.

More concrete and definite values for $\Delta ^{\!\!\mathcal{O}}$ are
obtainable when we exploit the knowledge about the underlying conformal
field theory more deeply. Considering an operator which in the conformal
limit corresponds to a primary field we can of course compute the conformal
dimension by appealing to the ultraviolet limit of the two-point correlation
function 
\begin{equation}
\left\langle \mathcal{O}_{i}(r)\mathcal{O}_{j}(0)\right\rangle
=\sum_{k}C_{ijk}\,r^{2\Delta _{k}-2\Delta _{i}-2\Delta _{j}}\,\left\langle 
\mathcal{O}_{k}(0)\right\rangle +\ldots  \label{ultra1}
\end{equation}
The three-point couplings $C_{ijk}$ are independent of $r$. In particular
when assuming that $0$ is the smallest conformal dimension occurring in the
model (which is the case for unitary models), we have 
\begin{equation}
\lim_{r\rightarrow 0}\left\langle \mathcal{O}(r)\mathcal{O}(0)\right\rangle
\sim r^{-4\Delta ^{\!\!\mathcal{O}}}\,\,\qquad \qquad \text{for }r\ll \left( 
\frac{C_{\Delta ^{\mathcal{O}}\Delta ^{\mathcal{O}}\,0}}{C_{\Delta ^{%
\mathcal{O}}\Delta ^{\mathcal{O}}\Delta ^{\mathcal{O}^{\prime \prime
}}}\left\langle \mathcal{O}^{^{\prime \prime }}\right\rangle }\right)
^{1/2\Delta ^{\mathcal{O}^{\prime \prime }}}\,.  \label{ultra}
\end{equation}
Here $\mathcal{O}^{\prime \prime }$ is the operator with the second smallest
dimension for which the vacuum expectation value is non-vanishing. Using a
Lorentz transformation to shift the $\mathcal{O}(r)$ to the origin and
expanding the correlation function in terms of form factors in the usual
fashion 
\begin{eqnarray}
\left\langle \mathcal{O}(r)\mathcal{O}^{\prime }(0)\right\rangle
&=&\sum_{n=1}^{\infty }\sum_{\mu _{1}\ldots \mu _{n}}\int\limits_{-\infty
}^{\infty }\ldots \int\limits_{-\infty }^{\infty }\frac{d\theta _{1}\ldots
d\theta _{n}}{n!(2\pi )^{n}}\exp \left( -r\sum_{i=1}^{n}m_{\mu _{i}}\cosh
\theta _{i}\right)  \nonumber \\
&&\times \,\,F_{n}^{\mathcal{O}|\mu _{1}\ldots \mu _{n}}(\theta _{1},\ldots
,\theta _{n})\,\left( F_{n}^{\mathcal{O}^{\prime }|\mu _{1}\ldots \mu
_{n}}(\theta _{1},\ldots ,\theta _{n})\,\right) ^{*}.  \label{corr}
\end{eqnarray}
we can compute the l.h.s. of (\ref{ultra}) and extract $\Delta ^{\!\!%
\mathcal{O}}$ thereafter. The disadvantage to proceed in this way is
many-fold. First we need to compute the multidimensional integrals in (\ref
{corr}) for each value of $r$, which means to produce a proper curve
requires a lot of computational (at present computer) time. Second we need
already a relatively good guess for $\Delta ^{\!\!\mathcal{O}}$. Third for
very small $r$ the $n$-th term within the sum is proportional to ($\log
(r))^{n}$ such that we have to include more and more terms in that region
and fourth we need the precise values of the lowest non-vanishing form
factors, i.e. in general vacuum expectation values or one particle form
factors to compute the r.h.s. of (\ref{corr}). However, the lowest
non-vanishing form factor can be of an arbitrary particle number and one may
still extract the value of $\Delta ^{\!\!\mathcal{O}}$.

A short remark is also due concerning solutions related to different sets of 
$\mu $'s. The sum over the particle types simplifies considerably when
taking into account that form factors corresponding to two sets, which
differ only by a permutation, lead to the same contribution in the sum. This
follows simply by using one of Watson's equations \cite
{Kar,Smir,Zamocorr,BFKZ}, which states that when two particles are
interchanged we will pick up the related two particle scattering matrix as a
factor. Noting that the scattering matrix is a phase, the expression remains
unchanged.

\subsection{$\Delta $-sum rules}

Most of the disadvantages, which emerge when using (\ref{ultra}) to compute
the conformal dimensions, can be circumvented by formulating sum rules in
which the $r$-dependence has been eliminated. Such type of rule has for
instance been formulated by Smirnov \cite{clust} already more than a decade
ago. However, the rule stated there is slightly cumbersome in its evaluation
and we will therefore resort to one found more recently by Delfino,
Simonetti and Cardy \cite{DSC}. In close analogy to the spirit and
derivation of the c-theorem \cite{ZamC} these authors derived an expression
for the difference between the ultraviolet and infrared conformal dimension
of a primary field $\mathcal{O}$%
\begin{equation}
\Delta _{uv}^{\!\!\mathcal{O}}-\Delta _{ir}^{\!\!\mathcal{O}}=-\frac{1}{%
2\left\langle \mathcal{O}\right\rangle }\int\limits_{0}^{\infty
}r\left\langle \Theta (r)\mathcal{O}(0)\right\rangle \,dr\,\,.
\label{deldel}
\end{equation}
Using the expansion of the correlation function in terms of form factors (%
\ref{corr}) we may carry out the$\ r$-integration in (\ref{deldel}) and
obtain 
\begin{eqnarray}
\Delta _{uv}^{\!\!\mathcal{O}}-\Delta _{ir}^{\!\!\mathcal{O}} &=&-\frac{1}{%
2\left\langle \mathcal{O}\right\rangle }\sum_{n=1}^{\infty }\sum_{\mu
_{1}\ldots \mu _{n}}\int\limits_{-\infty }^{\infty }\ldots
\int\limits_{-\infty }^{\infty }\frac{d\theta _{1}\ldots d\theta _{n}}{%
n!(2\pi )^{n}\left( \sum_{i=1}^{n}m_{\mu _{i}}\cosh \theta _{i}\right) ^{2}}
\nonumber \\
&&\times F_{n}^{\Theta |\mu _{1}\ldots \mu _{n}}(\theta _{1},\ldots ,\theta
_{n})\,\left( F_{n}^{\mathcal{O}|\mu _{1}\ldots \mu _{n}}(\theta _{1},\ldots
,\theta _{n})\,\right) ^{*}\,\,\,.  \label{dcorr}
\end{eqnarray}
Notice also that unlike in the evaluation of the c-theorem, which deals with
a monotonically increasing series, due to the fact that it only involves
absolute values of form factors, the series (\ref{deldel}) can in principle
be alternating. Before the concrete evaluation of the expression (\ref{dcorr}%
) for the various solutions we constructed for the SU(3)$_{2}$-HSG model, we
should pause for a while and appreciate the advantages of this formula in
comparison with (\ref{ultra}). First of all, since the $r$-dependence has
been integrated out we only have to evaluate the multidimensional integrals
once. Second the evaluation of (\ref{dcorr}) does not involve any
anticipation of the value of $\Delta ^{\!\!\mathcal{O}}$. Third one is
very often in the comfortable position that despite the fact that 
the vacuum expectation value occurs explicitly in the sum rule, its
explicit form is not needed. Once it is non-vanishing, the next lowest 
non-vanishing form factor may be normalized such that
$\left\langle \mathcal{O}\right\rangle $ cancels from the whole
expression. For a singular vacuum expectation value the value of the
remaining integral guarantees that the sum rule maintains its form
as for instance discussed in \cite{dis}. Finally and most
important fourth, the difficulty to identify the suitable region in $r$
which is governed by the $(\log r)^{n}$ behaviour of the $n$-th term in the
sum in (\ref{corr}) and the upper bound in (\ref{ultra1}) has completely
disappeared.

There are however little drawbacks for theories with internal symmetries and
for the case when the lowest non-vanishing form factor of the operator we
are interested in is not the vacuum expectation value. The first problem
arises due to the fact that the sum rule is only applicable for primary
fields $\mathcal{O}$ whose two-point correlation function with the energy
momentum tensor is non-vanishing. Since in our model the $n$-particle form
factors related to the energy momentum tensor are only non-vanishing for
even particle numbers, we may only use it for the operators $\mathcal{O}%
_{0,0}^{0,0}$, $\mathcal{O}_{0,2}^{0,1}$, $\mathcal{O}_{2,0}^{1,0}$ and $%
\mathcal{O}_{2,2}^{1,1}$, where the latter operator is plagued be the second
problem.

We will now compute the sum rule for the operators $\mathcal{O}_{0,0}^{0,0}$%
, $\mathcal{O}_{0,2}^{0,1}$, $\mathcal{O}_{2,0}^{1,0}$ up to the 6-particle
contribution. We commence with the two particle contribution which is always
evaluated effortlessly. Noting that 
\begin{equation}
F_{2}^{\Theta }(\theta )=-2\pi im^{2}\sinh (\theta /2)\,
\end{equation}
and the fact that $\Delta _{ir}^{\!\!\mathcal{O}}$ is zero in a purely
massive model, the two particle contribution acquires a particular simple
form 
\begin{equation}
(\Delta ^{\!\!\mathcal{O}})^{(2)}=\frac{i}{4\pi \left\langle \mathcal{O}%
\right\rangle }\int\limits_{-\infty }^{\infty }d\theta \frac{\tanh \theta }{%
\cosh \theta }\left( F_{2}^{\mathcal{O}|++}(2\theta )\right) ^{\ast
}\,\,\,.\,  \label{twop}
\end{equation}
Using now the explicit expressions for the two-particle form factors (\ref
{twopp}), we immediately find 
\begin{equation}
(\Delta ^{\!\!\mathcal{O}_{0,0}^{0,0}})^{(2)}=(\Delta ^{\!\!\mathcal{O}%
_{0,2}^{0,1}})^{(2)}=(\Delta ^{\!\!\mathcal{O}_{2,0}^{1,0}})^{(2)}=1/8\,\,\,.
\end{equation}
We recall, see also \cite{CFK}, that in the limit $\sigma \rightarrow \infty 
$ we obtain two copies of the thermally perturbed Ising model. This means
that in the sum over the particle types in (\ref{dcorr}) there will be no
contributions from terms involving different types of particles. We only
obtain two equal contributions, namely $1/16$ from  $F_{2}^{\mathcal{O}|++}$
and $F_{2}^{\mathcal{O}|--}$, such that  the operator $\mathcal{O}%
_{0,0}^{0,0}$ plays the role of the disorder operator, as we expect.

To distinguish the operators $\mathcal{O}_{0,0}^{0,0}$, $%
\mathcal{O}_{0,2}^{0,1}$, $\mathcal{O}_{2,0}^{1,0}$ from each other we have
to proceed to higher particle contributions. At present there exist no
analytical arguments for this and we therefore resort to a brute force
numerical computation.

\noindent Denoting by $(\Delta ^{\!\!\mathcal{O}})^{(n)}$ the contribution
up to the $n$-th particle form factor, our numerical Monte Carlo integration%
\footnote{%
We employed here the widely used numerical recipe routine VEGAS \cite{NUM}.
Typical standard deviations we achieve correspond to the order of the last
digit we quote.} yields 
\begin{eqnarray}
\qquad (\Delta ^{\!\!\mathcal{O}_{0,0}^{0,0}})^{(4)} &=&0.0987\qquad (\Delta
^{\!\!\mathcal{O}_{0,0}^{0,0}})^{(6)}=0.1004,  \label{98} \\
\qquad (\Delta ^{\!\!\mathcal{O}_{0,2}^{0,1}})^{(4)} &=&0.0880\qquad (\Delta
^{\!\!\mathcal{O}_{0,2}^{0,1}})^{(6)}=0.0895, \\
\qquad (\Delta ^{\!\!\mathcal{O}_{2,0}^{1,0}})^{(4)} &=&0.0880\qquad (\Delta
^{\!\!\mathcal{O}_{2,0}^{1,0}})^{(6)}=0.0895\,\,.  \label{99}
\end{eqnarray}

\noindent We shall be content with the precision reached at this point, but
we will have a look at the overall sign of the next contribution. From the
explicit expressions of the 8-particle form factors we see that for $%
\mathcal{O}_{0,0}^{0,0}$ the next contribution will reduce the value for $%
\Delta $. For the other two operators we have several contributions with
different signs, such that the overall value is not clear a priory. In this
light, we conclude that the operators $\mathcal{O}_{0,0}^{0,0}$, $\mathcal{O}%
_{0,2}^{0,1}$, $\mathcal{O}_{2,0}^{1,0}$ all possess conformal dimension $%
1/10$ in the ultraviolet limit. Unfortunately, the values for the latter 
two operators do not allow such a clear cut deduction as for the first one. 
Nonetheless, we base our statement on the knowledge of the operator
content of the conformal field theory and confirm them also by elaborating
directly on (\ref{ultra}) and (\ref{corr}).

\begin{center}
\includegraphics[width=12.65cm,height=9.3cm]{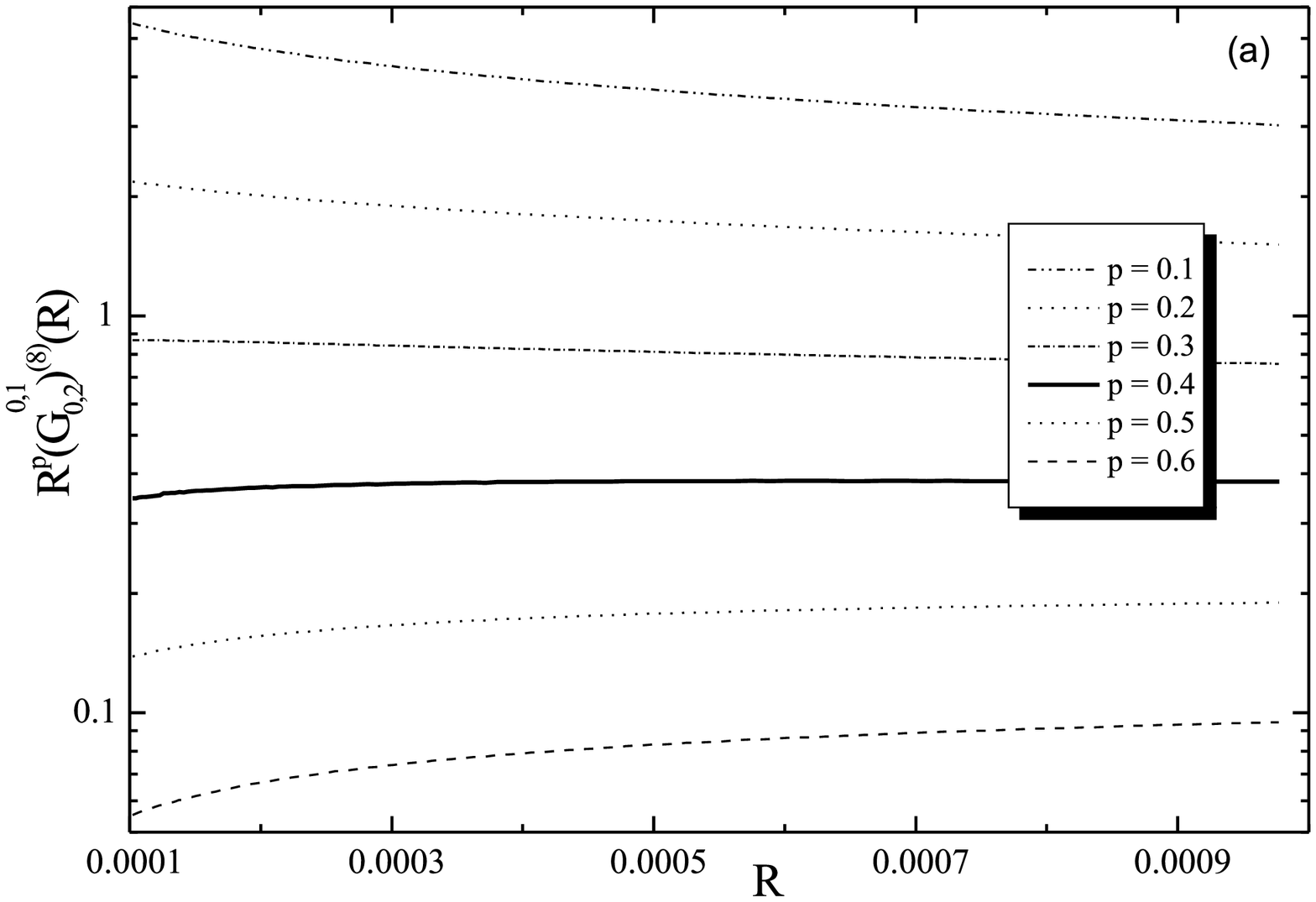}
\end{center}

\begin{center}
\includegraphics[width=9.3cm,height=12.65cm,angle=-90]{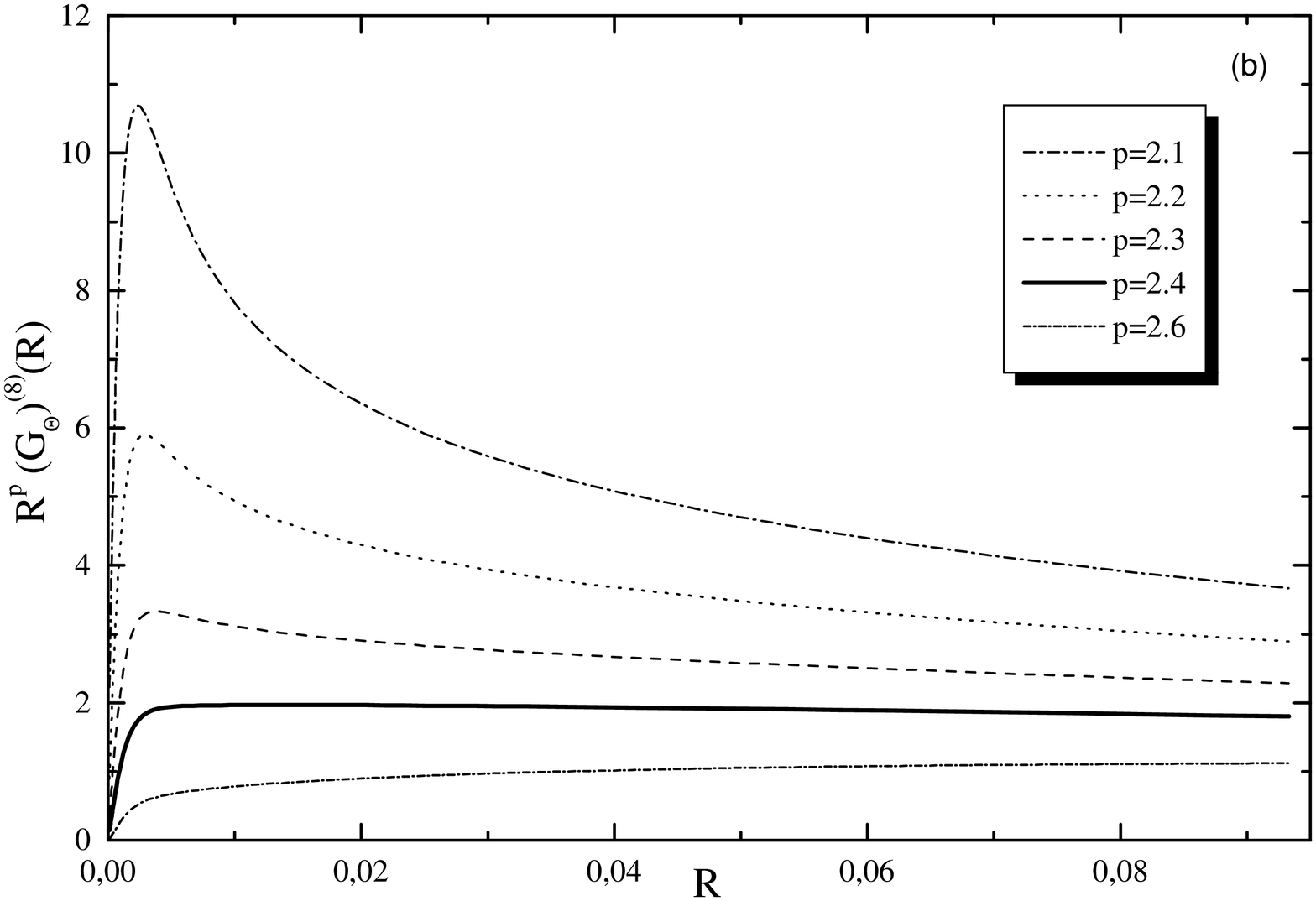}
\end{center}

\noindent {\small Figure 2: Rescaled correlation function $%
G_{0,2}^{0,1}(R):=\left\langle \mathcal{O}_{0,2}^{0,1}(R)\mathcal{O}%
_{0,2}^{0,1}(0)\right\rangle $ part (a) and 
($G_{\Theta})^{(8)}(R):=\left\langle \Theta (R)\Theta (0)\right\rangle $ 
part (b)
summed up to the eight particle contribution
as a function of $R=rm$.}

\subsection{$\Delta $ from correlation functions}

First of all we do not presume anything about the conformal dimension of the
operator $\mathcal{O}$ and multiply its two-point correlation function (\ref
{corr}) by $r^{p}$ with $p$ being some arbitrary power. Once this
combination behaves as a constant in the vicinity of $r=0$ we take this
value as the first non-vanishing three-point coupling divided by the vacuum
expectation value of $\mathcal{O}$ and $p/4\,$as its conformal dimension.
This means even without knowing the vacuum expectation value we have a
rational to fix $p$, but we can not determine the first term in (\ref{ultra1}%
). Figure 2a) exhibits this analysis for the operator $\mathcal{O}_{0,2}^{0,1}$
up to the $8$-particle contribution and we conclude from there that its
conformal dimension is $1/10$. For the other operators the figures look
qualitatively the same.

The results of the same type of analysis for the energy momentum tensor is
depicted in figure 2b), from which we deduce the conformal dimension $3/5$.
Recalling that the energy-momentum tensor is proportional \cite{Cardy} to
the dimension of the perturbing field this is precisely what we expected to
find.

Furthermore, we observe that the relevant interval for $r$ differs by two
orders of magnitude, which by taking the upper bound for the validity of (%
\ref{ultra}) into account should amount to $C_{\frac{1}{10}\frac{1}{10}0}C_{%
\frac{3}{5}\frac{3}{5}\frac{1}{10}}/(C_{\frac{3}{5}\frac{3}{5}0}C_{\frac{1}{%
10}\frac{1}{10}\frac{1}{10}})\sim \mathcal{O}(10^{-2})$. Since to our
knowledge these quantities have not been computed from the conformal side,
this inequality can not be double checked at this stage.



\begin{center}
\includegraphics[width=9.4cm,height=12.65cm,angle=-90]{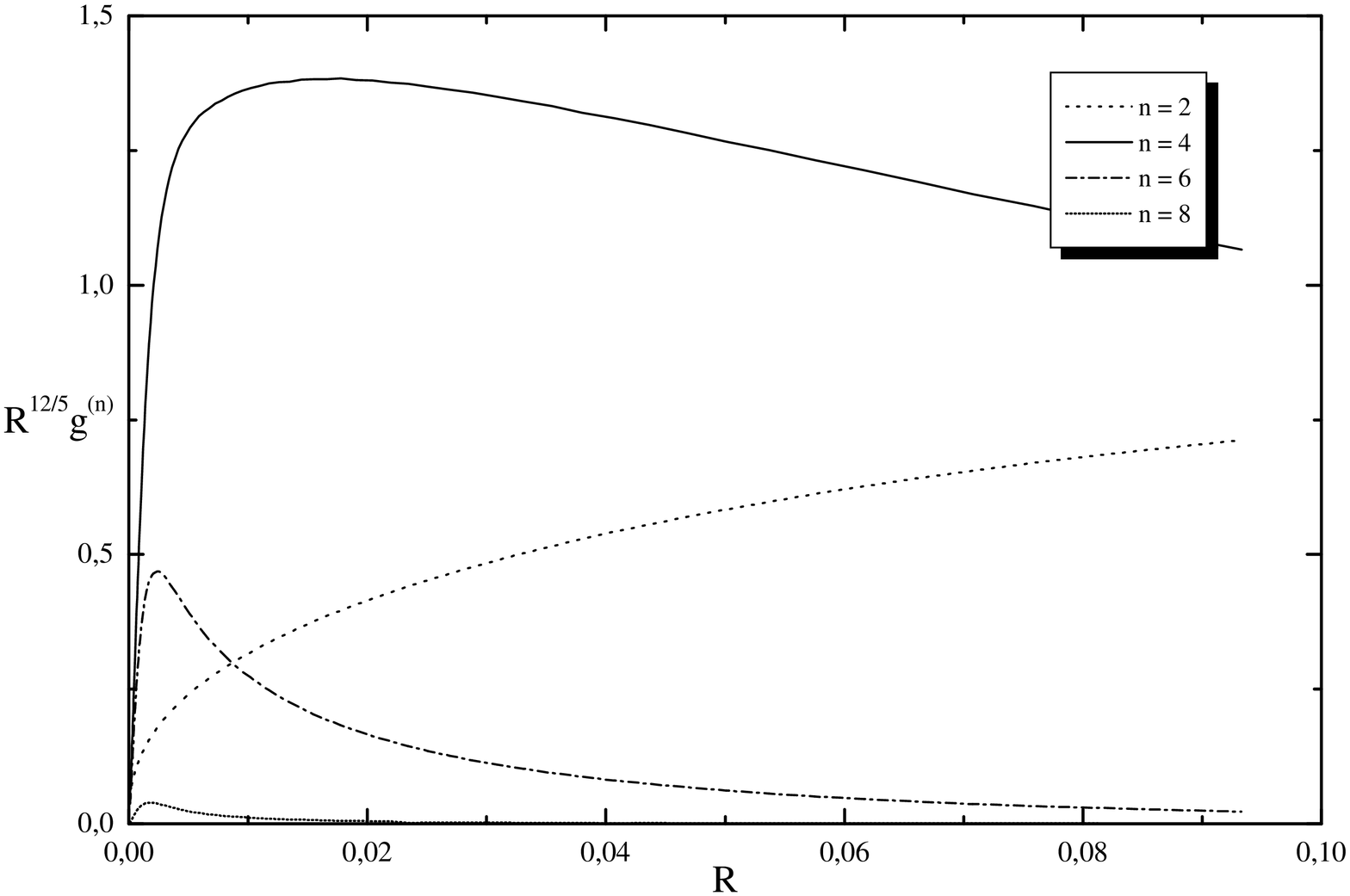}
\end{center}

\noindent {\small Figure 3: Rescaled individual }$n${\small -particle
contribution $g^{(n)}(R)$ to the correlation function.}

In figure 3 we also exhibit the individual $n$-particle contributions.
Excluding the two particle contribution, these data also confirm the
proportionality of the $n$-th term to ($\log (r))^{n}$.

We have carried out similar analysis for the other solutions we have
constructed and report our findings in table 1. We observe that the
combination of the vacuum expectation value times the three-point coupling
for these operators differ, which is the prerequisite for unraveling the
degeneracy.

\section{Conclusions}

With regard to the main conceptual question addressed in this paper, we draw
the overall conclusion that solutions of the form factor consistency
equations can be identified with operators in the underlying ultraviolet
conformal field theory. In this sense one can give meaning to the operator
content of the integrable massive model. The quantity on which the
identification is based is the conformal dimension of the operator.
Naturally this implies that once the conformal field theory is degenerate in
this quantity, as it is the case for the model we investigated, the
identification can not be carried out in a one-to-one fashion and therefore
the procedure has to be refined. In principle this would be possible by
including the knowledge of the three-point coupling of the conformal field
theory and the vacuum expectation value into the analysis. The former
quantities are in principle accessible by working out explicitly the
conformal fusion structure, whereas the computation of the latter still
remains an open challenge. In fact what one would like to achieve ultimately
is the identification of the conformal fusion structure within the massive
models.

It would be desirable to put further constraints on the solutions by means
of other arguments, that is exploiting the symmetries of the model,
formulating quantum equations of motion, possibly performing perturbation
theory etc.

Technically we have confirmed that the sum rule (\ref{deldel}) is clearly
superior to the direct analysis of the correlation function. It would
therefore be highly desirable to develop arguments which also apply for
theories with internal symmetries and possibly to resolve the mentioned
degeneracies in the conformal dimensions.

It remains also an open question, whether the general solution procedure
presented in this manuscript can be generalized to the degree that the type
of determinants presented will serve as generic building blocks of form
factors.

The specific conclusions for the SU(3)$_{2}$-homogeneous Sine-Gordon model
are as follows: We have provided a rigorous proof for the solutions of the
form factor consistency equations which were previously stated in \cite{CFK}%
. In addition we found a huge number of new solutions. By means of the sum
rule and a direct analysis of the correlation functions we identified the
conformal dimension of these operators in the underlying conformal field
theory. Considering the total number of operators present in the conformal
field theory (see table 2) one still expects to find additional solutions,
in particular the identification of the fields possessing conformal
dimension $1/2$ is outstanding. Nonetheless, concerning the physical picture
presented for this model one can surely claim that it rests now on quite
firm ground. After the central charge of the conformal field theory had been
reproduced by means of the thermodynamic Bethe ansatz \cite{CFKM} and the
c-theorem in the context of the form factor program \cite{CFK}, we have now
also identified the dimension of various operators. In particular the
dimension of the perturbing operator was identified to be $3/5$.

\noindent \textbf{Acknowledgments: } A.F. is grateful to the Deutsche
Forschungsgemeinschaft (Sfb288) for financial support. O.A.C. thanks CICYT
(AEN99-0589), DGICYT (PB96-0960), and the EC Commission (TMR grant
FMRX-CT96-0012) for partial financial support and is also very grateful to
the Institut f\"{u}r theoretische Physik of the Freie Universit\"{a}t for
hospitality. We would like to thank J.L. Miramontes and R.A. V\'{a}zquez for
helpful discussions.

\section{Appendix}

\subsection{Elementary symmetric polynomials}

In this appendix we assemble several properties of elementary symmetric
polynomials to which we wish to appeal from time to time. Most of them may
be found either in \cite{Don} or can be derived effortlessly. The elementary
symmetric polynomials are defined as 
\begin{equation}
\sigma _{k}(x_{1},\ldots ,x_{n})=\sum_{l_{1}<\ldots <l_{k}}x_{l_{1}}\ldots
x_{l_{k}}\,\,\,.
\end{equation}
They are generated by 
\begin{equation}
\prod_{k=1}^{n}(x+x_{k})=\sum_{k=0}^{n}x^{n-k}\sigma _{k}(x_{1},\ldots
,x_{n})\,\,.
\end{equation}
and as a consequence may also be represented in terms of an integral
representation 
\begin{equation}
\text{\thinspace }\sigma _{k}(x_{1},\ldots ,x_{n})\,=\frac{1}{2\pi i}%
\oint_{|z|=\varrho }\frac{dz}{z^{n-k+1}}\prod\limits_{k=1}^{n}(z+x_{k})\,,
\label{symint}
\end{equation}
which is convenient for various applications. Here $\varrho $ is an
arbitrary positive real number.

\noindent With the help of (\ref{symint}) we easily derive the identity 
\begin{equation}
\sigma _{k}(-x,x,x_{1},\ldots ,x_{n})=\sigma _{k}(x_{1},\ldots
,x_{n})-x^{2}\sigma _{k-2}(x_{1},\ldots ,x_{n})\,\,,  \label{symp}
\end{equation}
which will be central for us. We will also require the asymptotic behaviours 
\begin{equation}
\mathcal{T}_{1,\eta }^{\lambda }\,\sigma _{k}(x_{1},\ldots ,x_{n})\sim
\QATOPD\{ . {e^{\eta \lambda }\sigma _{\eta }(x_{1},\ldots ,x_{\eta })\sigma
_{k-\eta }(x_{\eta +1},\ldots ,x_{n})\,\,\,\quad \text{for }\eta
<k}{e^{k\lambda }\sigma _{k}(x_{1},\ldots ,x_{\eta
})\,\,\,\,\,\,\,\,\,\,\,\,\,\,\,\,\,\,\,\,\,\,\,\,\,\,\,\,\,\,\,\,\,\,\,\,\,%
\,\quad \quad \quad \text{for }\eta \geq k}  \label{as+}
\end{equation}
and 
\begin{equation}
\mathcal{T}_{1,\eta }^{-\lambda }\,\sigma _{k}(x_{1},\ldots ,x_{n})\sim
\QATOPD\{ . {\sigma _{k}(x_{\eta +1},\ldots ,x_{n})\,\,\,\quad \qquad \qquad
\quad \text{for }\eta \leq n-k}{\frac{\sigma _{k+\eta -n}(x_{1},\ldots
,x_{\eta })\,\sigma _{n-\eta }(x_{\eta +1},\ldots ,x_{n})}{e^{\lambda
(k+\eta -n)}}\,\,\,\,\,\,\,\,\,\,\,\,\,\,\text{for }\eta >n-k}  \label{as-}
\end{equation}
which may be obtained from (\ref{symint}) as well.

\subsection{Explicit form factor formulae}

Having constructed the general solutions in terms of the parameterization (%
\ref{fact}), it is simply a matter of collecting all the factors to get
explicit formulae. For the concrete computation of the correlation function,
it is convenient to have some of the evaluated expressions at hand in form
of hyperbolic functions.

\subsubsection{One particle form factors}

\begin{equation}
F_{1}^{{\mathcal{O}}_{1,0}^{0,0}|+}=F_{1}^{{\mathcal{O}}%
_{0,1}^{0,0}|-}=F_{1}^{{\mathcal{O}}_{0,1}^{0,1}|-}=F_{1}^{{\mathcal{O}}%
_{1,0}^{1,0}|+}=H^{1,0}=H^{0,1}
\end{equation}

\subsubsection{Two particle form factors}

\begin{eqnarray}
F_{2}^{^{\!\!\mathcal{O}|\pm \pm }} &=&i\left\langle \mathcal{O}%
\right\rangle \tanh \tfrac{\theta }{2},\quad \qquad \text{for\quad }\mathcal{%
O}=\text{ }\mathcal{O}_{0,0}^{0,0}\text{, }\mathcal{O}_{0,2}^{0,1},\mathcal{O%
}_{2,0}^{1,0},\text{ }  \label{twopp} \\
F_{2}^{{\mathcal{O}}_{1,1}^{0,1}|+-} &=&H^{1,1}e^{\theta _{21}/2}F_{\text{min%
}}^{+-}(\theta ),\quad \quad F_{2}^{{\mathcal{O}}_{1,1}^{1,0}|+-}=H^{1,1}F_{%
\text{min}}^{+-}(\theta )
\end{eqnarray}

\subsubsection{Three particle form factors}

\begin{equation}
F_{3}^{{\mathcal{O}}|\pm \pm \pm }=\frac{H^{0,1}\prod_{i<j}F_{\text{min}%
}^{\mu _{i}\mu _{j}}(\theta _{ij})}{\prod_{1\leq i<j\leq 3}\cosh (\theta
_{ij}/2)}\quad \text{for\quad }{\mathcal{O}}_{1,0}^{0,0}\text{,}{\mathcal{O}}%
_{0,1}^{0,0}\text{, }{\mathcal{O}}_{0,1}^{0,1}\text{, }{\mathcal{O}}%
_{1,0}^{1,0}
\end{equation}

\subsubsection{Four particle form factors}

\begin{eqnarray}
F_{4}^{\Theta |++--} &=&\frac{-\pi m^{2}(2+\sum_{i<j}\cosh (\theta _{ij}))}{%
2\cosh (\theta _{12}/2)\cosh (\theta _{34}/2)}\prod_{i<j}\tilde{F}_{\text{min%
}}^{\mu _{i}\mu _{j}}(\theta _{ij}) \\
F_{4}^{\mathcal{O}_{0,0}^{0,0}|++--} &=&\frac{-\left\langle \mathcal{O}%
_{0,0}^{0,0}\right\rangle \cosh (\theta _{13}/2+\theta _{24}/2)}{2\cosh
(\theta _{12}/2)\cosh (\theta _{34}/2)}\prod_{i<j}\tilde{F}_{\text{min}%
}^{\mu _{i}\mu _{j}}(\theta _{ij})
\end{eqnarray}

\subsubsection{Five particle form factors}

\begin{equation}
F_{5}^{{\mathcal{O}}|\pm \pm \pm \pm \pm }=\frac{H^{0,1}\prod_{i<j}F_{\text{%
min}}^{\mu _{i}\mu _{j}}(\theta _{ij})}{\prod_{1\leq i<j\leq 5}\cosh (\theta
_{ij}/2)}\quad \text{for }{\mathcal{O}}_{1,0}^{0,0}\text{, }{\mathcal{O}}%
_{0,1}^{0,0}\text{, }{\mathcal{O}}_{1,0}^{1,0}\text{, }{\mathcal{O}}%
_{0,1}^{0,1}
\end{equation}

\subsubsection{Six particle form factors}

\begin{eqnarray}
F_{6}^{\Theta |++++--} &=&\frac{\pi m^{2}(3+\sum_{i<j}\cosh (\theta _{ij}))}{%
4\prod_{1\leq i<j\leq 4}\cosh (\theta _{ij}/2)}\prod_{i<j}\tilde{F}_{\text{%
min}}^{\mu _{i}\mu _{j}}(\theta _{ij}) \\
F_{6}^{\mathcal{O}_{0,0}^{0,0}|++++--} &=&\frac{\left\langle \mathcal{O}%
_{0,0}^{0,0}\right\rangle \left( (\sigma _{2}^{-})^{2}+\sigma
_{4}^{+}+\sigma _{2}^{+}\sigma _{2}^{-}/(\sigma _{2}^{+}+\sigma_2^- )
\right) }{16\cosh
(\theta _{56}/2)\prod_{1\leq i<j\leq 4}\cosh (\theta _{ij}/2)}\prod_{i<j}F_{%
\text{min}}^{\mu _{i}\mu _{j}}(\theta _{ij})
\end{eqnarray}

\subsubsection{Seven particle form factors}

\begin{equation}
F_{7}^{{\mathcal{O}}|\pm \pm \pm \pm \pm \pm \pm }=\frac{H^{0,1}%
\prod_{i<j}F_{\text{min}}^{\mu _{i}\mu _{j}}(\theta _{ij})}{\,\prod_{1\leq
i<j\leq 7}\text{cosh}(\theta _{ij}/2)}\quad \text{for }{\mathcal{O}}%
_{1,0}^{0,0}\text{, }{\mathcal{O}}_{0,1}^{0,0}\text{, }{\mathcal{O}}%
_{1,0}^{1,0}\text{, }{\mathcal{O}}_{0,1}^{0,1}
\end{equation}

\subsubsection{Eight particle form factors}

\begin{equation}
F_{8}^{\Theta |++------}=\frac{-\pi m^{2}(4+\sum_{i<j}\text{cosh}(\theta
_{ij}))\,\text{cosh}(\theta _{12}/2)}{8\,\prod_{3\leq i<j\leq 8}\text{cosh}%
(\theta _{ij}/2)}\prod_{i<j}\tilde{F}_{\text{min}}^{\mu _{i}\mu _{j}}(\theta
_{ij})
\end{equation}

\begin{equation}
F_{8}^{\Theta |++++----}=\frac{\pi m^{2}(\sigma _{4}^{-})^{1/2}(\sigma
_{1}^{-}\sigma _{3}^{+}+\sigma _{1}^{+}\sigma _{3}^{-})(4+\sum_{i<j}\text{%
cosh}(\theta _{ij}))}{2^{7}\,(\sigma _{4}^{+})^{3/2}\prod_{1\leq i<j\leq 4}%
\text{cosh}(\theta _{ij}/2)\prod_{5\leq i<j\leq 8}\text{cosh}(\theta _{ij}/2)%
}\prod_{i<j}F_{\text{min}}^{\mu _{i}\mu _{j}}(\theta _{ij})
\end{equation}

\begin{equation}
F_{8}^{\Theta |++++++--}=\frac{-\pi m^{2}(4+\sum_{i<j}\text{cosh}(\theta
_{ij}))\,\text{cosh}(\theta _{78}/2)}{8\,\prod_{1\leq i<j\leq 6}\text{cosh}%
(\theta _{ij}/2)}\prod_{i<j}\tilde{F}_{\text{min}}^{\mu _{i}\mu _{j}}(\theta
_{ij})
\end{equation}

\begin{description}
\item  {\small \setlength{\baselineskip}{12pt}}
\end{description}

\end{document}